\documentclass[a4paper]{article}
\usepackage[utf8]{inputenc}  
\usepackage[T1]{fontenc}
\usepackage{hyperref}
\usepackage[numbers]{natbib}
\usepackage{amsmath}
\usepackage{comment}
\usepackage{array}     
\usepackage{enumitem}  
\usepackage{graphicx}
\usepackage{subcaption}
\usepackage{xcolor}
\usepackage{booktabs}
\usepackage{float}
\usepackage[a4paper, margin=2.5cm]{geometry}
\usepackage{authblk}

\begin{document}
\let\WriteBookmarks\relax
\def\floatpagepagefraction{1}
\def\textpagefraction{.001}

\author[1,2,3]{Samuel Del~Fr\'e}
\author[1,2]{Andr\'ee De~Backer}
\author[2,3]{Christophe Domain}
\author[1,2]{Ludovic Thuinet}
\author[1,2]{Charlotte S. Becquart}

\affil[1]{Univ. Lille, CNRS, INRAE, Centrale Lille, UMR 8207 - UMET - Unit\'e Matériaux et Transformations, F-59000 Lille, France}

\affil[2]{EM2VM, Joint laboratory Study and Modelling of the Microstructure for Ageing of Materials, France}

\affil[3]{EDF-R\&D, D\'epartement Mat\'eriaux et M\'ecanique des Composants, Les Renardi\`eres, F-77250, Moret sur Loing, France}

\affil[3]{Univ. Lille, CNRS, UMR 8523 – PhLAM – Physique des Lasers Atomes et Molécules, F-59000 Lille, France}





\date{}

\title{Unsupervised Machine-Learning Pipeline for Data-Driven Defect Detection and Characterisation: Application to Displacement Cascades}

\maketitle

\begin{figure}[H]
	\centering
	\includegraphics[width=\textwidth]{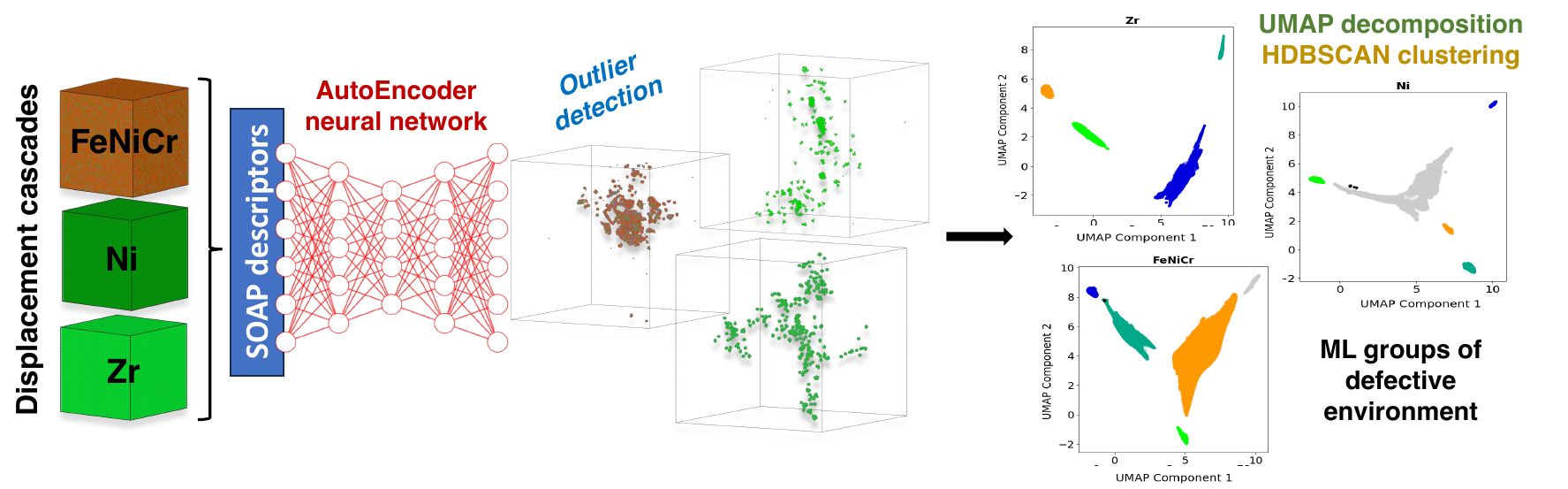}
\end{figure}

\begin{abstract}
	Neutron irradiation produces, within a few picoseconds, displacement cascades that are sequences of atomic collisions generating point and extended defects which subsequently affects the long-term evolution of materials. The diversity of these defects, characterized morphologically and statistically, defines what is called the "primary damage". In this work, we present a fully unsupervised machine learning (ML) workflow that detects and classifies these defects directly from molecular dynamics data. Local environments are encoded by the Smooth Overlap of Atomic Positions (SOAP) vector, anomalous atoms are isolated with autoencoder neural networks (AE), embedded with Uniform Manifold Approximation and Projection (UMAP) and clustered using Hierarchical Density‑Based Spatial Clustering of Applications with Noise (HDBSCAN). Applied to 80~keV displacement cascades in Ni, Fe\textsubscript{70}Ni\textsubscript{10}Cr\textsubscript{20}, and Zr, the AE successfully identify the small fraction of outlier atoms that participate in defect formation. HDBSCAN then partitions the UMAP latent space of AE‑flagged SOAP descriptors into well defined groups representing vacancy‑ and interstitial‑dominated regions and, within each, separates small from large aggregates, assigning 99.7~\% of outliers to compact physical motifs. A signed cluster‑identification score confirms this separation, and cluster size scales with net defect counts (\(R^{2}>0.89\)). Statistical cross analyses between the ML outlier map and several conventional detectors (centrosymmetry, dislocation extraction, etc.) reveal strong overlap and complementary coverage, all achieved without template or threshold tuning. This ML workflow thus provides an efficient tool for the quantitative mapping of structural anomalies in materials, particularly those arising from irradiation damage in displacement cascades.
\end{abstract}

\section{Introduction}

Radiation-induced damage in structural materials is a critical challenge for applications such as nuclear reactors, as it directly compromises the structural integrity and long-term performance of the materials themselves \cite{liuReviewSynergisticDamage2024,pomaroReviewRadiationDamage2016,zinkleStructuralMaterialsFission2009}. When energetic particles (e.g., neutrons) collide with atoms in a crystal lattice, they produce primary knock-on atoms (PKAs) that in turn trigger collision cascades, generating a variety of point defects (vacancies and interstitials), extended defects (stacking faults, dislocation loops), and complex clusters \cite{brinkmanProductionAtomicDisplacements1956,NORDLUND2018450}. The evolution of these defects leads to macroscopic phenomena such as hardening, embrittlement, swelling and creep \cite{xiaoFundamentalMechanismsIrradiationHardening2019,realiAtomisticSimulationsAthermal2025}, yet their atomistic origins remain challenging to characterize experimentally, especially during the earliest femto- to picosecond timescales of cascade formation. 
The so-called primary damage created during cascade events defines the initial population of defects from which all subsequent microstructural evolution emerges \cite{nordlundHistoricalReviewComputer2019, stoller111PrimaryRadiation}. Over reactor-relevant timescales (years to decades), these defects migrate, interact, and transform, ultimately governing the stability of the microstructure and thus the macroscopic mechanical properties of structural materials. Since direct experimental access to atomistic defect dynamics at such extended timescales is not possible, predictive modeling frameworks must instead rely on an accurate description of the types, numbers,  configurations and properties of defects generated at the cascade stage, which then serve as input for higher-scale simulations of long-term material behavior.
Classical molecular dynamics (MD) simulations offer an atomistic framework to model such displacement cascades, enabling detailed observation of defect formation and evolution over picoseconds timescales \cite{becquartAtomisticModelingRadiation2018, becquartModelingMicrostructureIrradiation2011}. 
While these simulations offer valuable insights into the dynamics of defect formation in various materials at different PKA energies, identifying atoms that differ from the bulk at the end of cascades (that is, those associated with defects) is not straightforward. Most conventional defect identification methods rely on analyzing the local bonding neighborhood around each atom and comparing it to known specific topological patterns. Among the methods used are Centrosymmetry Parameter (CS) \cite{kelchnerDislocationNucleationDefect1998}, Common Neighbor Analysis (CNA) \cite{honeycuttMolecularDynamicsStudy1987,fakenSystematicAnalysisLocal1994}, topological cluster identification \cite{malinsIdentificationStructureCondensed2013}, Polyhederal Template Matching (PTM) \cite{larsenRobustStructuralIdentification2016}, Dislocation Analysis (DXA) \cite{stukowskiExtractingDislocationsNondislocation2010,stukowskiAutomatedIdentificationIndexing2012}, Voronoï cell analysis (VOR) \cite{lazarTopologicalFrameworkLocal2015a} and classical lattice site analysis \cite{debackerModellingPrimaryDamage2021}, though this is not an exhaustive list. These methods provide valuable insights into defect position and type across various applications. 
However, they may struggle to detect subtle defect features, especially under conditions such as thermal fluctuations and strain \cite{mollerBDANovelMethod2016,larsenRobustStructuralIdentification2016}, in the presence of complex defect structures \cite{snowSimpleApproachAtomic2019}, or when encountering defect types not represented in the method’s structural database \cite{furstossAllaroundLocalStructure2025}. They also impose challenges of CPU time, memory requirements, and scalability, as for instance neighbor searches, template matching, and Voronoi/topological routines grow quickly in cost with system and trajectory size, which constrains analyses unless parallelized or subsampled.
These limitations have driven the integration of machine learning (ML) and deep learning (DL) into atomistic materials simulations, significantly advancing the automatic detection and characterization of structural defects and crystal structures. These data-driven techniques typically operate by embedding local atomic environments (LAE) into high-dimensional descriptor spaces using representations such as Smooth Overlap of Atomic Positions (SOAP) and SO(4) bispectrum components \cite{bartokRepresentingChemicalEnvironments2013}, Steinhardt order parameters \cite{steinhardtBondorientationalOrderLiquids1983,lechnerAccurateDeterminationCrystal2008}, Behler-Parinello symmetry functions \cite{behlerGeneralizedNeuralNetworkRepresentation2007,behlerAtomcenteredSymmetryFunctions2011a} and graph-based encodings \cite {alleraNeighborsMapEfficient2024,banikCEGANNCrystalEdge2023}. Once encoded, a variety of algorithms ranging from statistical models to neural networks are used to detect, classify, or group atomic configurations. 
Unsupervised strategies based on statistical distance metrics allow the identification of atoms that deviate from reference defect-free distributions, enabling the detection of point and extended defects with high resolution \cite{goryaevaReinforcingMaterialsModelling2020,vontoussaintFaVADSoftwareWorkflow2021}. Probabilistic models such as Gaussian Mixture Models (GMMs) have been applied in both supervised and unsupervised modes to classify structural motifs or reveal hidden phases and nucleation pathways \cite{furstossAllaroundLocalStructure2025,beckerUnsupervisedTopologicalLearning2022}. 
Neural networks (NN), including feedforward architectures \cite {geigerNeuralNetworksLocal2013, chungDatacentricFrameworkCrystal2022}, convolutional neural networks (CNNs) \cite{alleraNeighborsMapEfficient2024,zilettiInsightfulClassificationCrystal2018}, Bayesian NN \cite{leithererRobustRecognitionExploratory2021} or autoencoders (AE) \cite{boattiniUnsupervisedLearningLocal2019,zhangFastCrystalGrowth2023}, have demonstrated strong performance in classifying crystalline structures, even under thermal noise or incomplete data. In particular, AE networks provide a highly flexible tool for unsupervised defect detection in atomistic simulations. These networks are trained to reconstruct their input data by learning a compressed internal representation (latent space) \cite{berahmandAutoencodersTheirApplications2024}. 
Therefore, by learning to reconstruct normal (defect-free) LAE, autoencoders tend to produce outliers with higher reconstruction errors when encountering structural anomalies, since these configurations deviate from the training distribution. In the context of atomistic simulations, this property enables the efficient and unsupervised identification of atoms associated with defect structures, without relying on labelled data.

Concerning displacement cascades \cite{bhardwajClassificationClustersCollision2020}, density-based clustering algorithms such as Hierarchical Density-Based Spatial Clustering of Applications with Noise (HDBSCAN) \cite{campelloHierarchicalDensityEstimates2015}, combined with non-linear dimensionality reduction techniques like Uniform Manifold Approximation and Projection (UMAP) \cite{mcinnesUMAPUniformManifold2018}, have enabled the unsupervised morphological classification of complex defect clusters. 
This UMAP + HDBSCAN combination has also been successfully used for characterization of local coordination environments within MD data \cite{roncoroniUnsupervisedLearningRepresentative2023,kyvalaUnsupervisedIdentificationCrystal2025}. In the following discussion, the term cluster will refer exclusively to physical defect clusters. To avoid any misunderstanding, clusters in the machine learning sense (i.e., those obtained with algorithms such as HDBSCAN) will be referred to as groups. UMAP algorithm is capable of preserving both local neighborhood relationships and global data structure when reducing dimensionality, making it superior to linear dimensionality reduction methods like Principal Component Analysis (PCA) \cite{mackiewiczPrincipalComponentsAnalysis1993} and more topology-preserving than t-distributed Stochastic Neighbor Embedding (t-SNE) \cite{JMLR:v9:vandermaaten08a}. This dual preservation ensures that intrinsically separable groups in the high-dimensional space remain well-defined in the low-dimensional embedding, often resulting in more compact and distinct groups \cite{allaouiConsiderablyImprovingClustering2020, herrmannEnhancingClusterAnalysis2024}. Building on this structure-preserving embedding, HDBSCAN automatically finds groups of arbitrary shapes and densities, constructs a hierarchical density landscape, and extracts stable, noise-resistant groups using persistence metrics without needing a predefined number of groups. The resulting UMAP~$\rightarrow{}$HDBSCAN pipeline thus delivers a robust, unsupervised framework optimized for identifying intricate, noisy groups structures making it ideal for complex defect-cluster classification tasks.

In this study, we combine the expressive power of SOAP descriptors for accurately encoding LAE with the anomaly detection capabilities of AE neural networks to identify atoms involved in structural defects in displacement cascades. In our framework, outliers are defined as atoms whose SOAP-based fingerprint vectors yield a reconstruction error above a chosen threshold when processed by the autoencoder. Such atoms are considered defective and are interpreted as participating in a physical defect. Once these outliers are detected, the resulting dataset is analyzed using UMAP for nonlinear dimensionality reduction, followed by HDBSCAN clustering to enable the unsupervised classification of structurally similar patterns into distinct defect types. 
This data-driven machine-learning framework, combining SOAP descriptors with autoencoders, UMAP reduction, and HDBSCAN clustering is applied to three sets of displacement cascade simulations in both face-centered cubic (fcc) and hexagonal-close packed (hcp) crystalline systems, namely Ni and FeNiCr from \cite{nairStatisticalStudyDisplacement2025} representative of austenitic stainless steels, and Zr materials. For a statistical comparison, the results obtained using the machine learning approach are evaluated against those obtained with some aforementioned conventional defect detection methods, including the classical lattice site analysis, CS, PTM, DXA, VOR and coordination analysis (CA). Specifically, we compare the number of atoms identified as outliers by each method, analyze recall and precision metrics between the ML framework and these conventional approaches, and provide a particular focus on the correspondence between ML-flagged groups and PTM-identified icosahedral (ICO) environments. The structure of this article is as follows: Section \ref{methods} describes the methods, including displacement cascade simulations, ML methodology, and conventional defect-characterisation techniques; Section \ref{full_results} presents and analyses the results, with emphasis on outlier detection, unsupervised defect classification, and comparison with reference approaches; Section \ref{conclusion} and \ref{perspectives} provides the main conclusions and perspectives for future work.

\section{Methods}\label{methods}
\subsection{Displacement Cascades modelling}\label{MD}

All classical molecular dynamics simulations were performed using the DYMOKA code \cite{DYMOKA}. The simulation box size was adapted according to the PKA energy to avoid any interaction between replicated image, as periodic boundary conditions were applied in all directions during the simulations. Since this study focuses exclusively on a PKA energy of 80~keV, the box dimensions were set to $98\times{}98\times{}98$ unit cells (3764768 atoms) for both FeNiCr and Ni fcc materials, and to $122\times{}152\times{}76$ unit cells (2818688 atoms) for hcp Zr. 
For the simulations, the empirical interaction potentials used were derived using the Embedded Atom Method (EAM) approach. The FeNiCr potential was originally derived by \cite{Bonny2011} and hardened by \cite{Beland2017}, the Ni potential was developed by \cite{Mishin2004} and hardened by \cite{Samolyuk2016}, and the Zr potential was developed and hardened by \cite{Mendelev2007}.
Before simulating displacement cascades, systems were thermalized using canonical MD at 100~K using velocity scaler as implemented in DYMOKA for 3~ps with time step of 1~fs. Once the system was equilibrated, displacement cascades were initiated in the microcanonical ensemble from different snapshots of the thermalization trajectory for up to 50~ps. A single atom was selected as the PKA and assigned a kinetic energy of 80~keV. The directions of the initial velocity were chosen to represent a statistically reasonable average behavior, following the approach described in \cite{stollerPrimaryDamageFormation1997}. The timestep was dynamically adjusted between 0.005~fs and 1~fs during the simulations to ensure accurate energy conservation throughout the cascade events. Primary damages were then analysed at the end of each simulation.
It should be noted that this classical MD-based computational protocol has already been applied to other systems such as Fe and W \cite{becquartModellingPrimaryDamage2021,debackerModellingPrimaryDamage2021}, and that a comprehensive description of the simulation results for Ni and FeNiCr is available in a recent publication \cite{nairStatisticalStudyDisplacement2025}. 
A more thorough investigation of Zr displacement cascades will be presented in an upcoming publication \cite{Debacker_zr2025}. 

\subsection{ML methodology}\label{ML}

\begin{figure}
	\centering
    \includegraphics[width=0.95\textwidth]{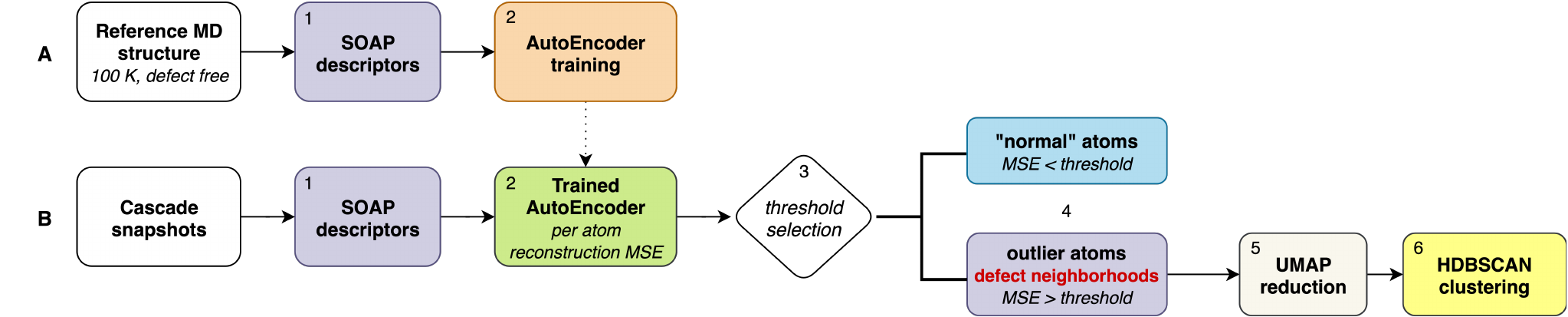}
	\caption{Schematic representation of the two branches  workflow (A and B) for ML-assisted defect mapping described in section \ref{ML}. SOAP descriptors are computed (A.1) for a defect-free MD reference structure (fcc or hcp depending the materials) to train an autoencoder neural network (A.2). For each cascade snapshot, SOAP descriptors are computed using the same parameters than the reference structure (B.1). Each SOAP vector then passed through the trained AE (B.2) for which a reconstruction error (MSE) threshold is selected (B.3) to classify atoms as inliers (MSE < threshold) or outliers forming defect neighbourhoods (MSE > threshold) based on the per-atom reconstruction error (B.4). Outlier atoms are then embedded with UMAP (B.5) and grouped with HDBSCAN (B.6) to yield unsupervised defect-type groups.}
	\label{Fig0_diag}
\end{figure}

A schematic representation of the machine learning pipeline is shown in Figure~\ref{Fig0_diag} and in the following discussion, the specific step of the workflow (A.1–3 and B.1–6) relevant to each result will be explicitly indicated.
The first step of the ML workflow (steps A.1 and B.1) involves encoding local atomic environments using high-resolution descriptors that are invariant to rotation, translation, and atomic indexing. These descriptors serve as input features for the subsequent stages of the machine learning analysis. In this work, SOAP descriptors was computed for each atom in the simulation box using DScribe \cite{dscribe,dscribe2} although the workflow is general and applicable to any descriptor that produces a vector-based atomic fingerprint. For fcc materials (Ni and FeNiCr), a cutoff radius of 4.7~Å was used, with $n_{\text{max}} = 4$ radial basis functions and a spherical harmonic degree of $l_{\text{max}} = 4$, resulting in 50-dimensional descriptor vectors. For the hcp material (Zr), a larger cutoff of 5.8~Å was employed, with $n_{\text{max}} = 6$ and $l_{\text{max}} = 4$, yielding 105-dimensional vectors. For both material types, the standard deviation $\sigma$ of the Gaussians used to expand the atomic density was fixed at 0.25. Note that the cut-off values were determined according to the radial distribution functions of both materials, in order to include first coordination shells within the local atomic environment descriptors. 
The second step (B.2) consists of outlier detection using autoencoder neural networks trained on SOAP fingerprints from defect-free reference structures (A.2). The reference SOAP matrix for fcc materials was constructed from a pure $21\times21\times21$ Ni simulation box equilibrated at 100~K using canonical MD (see Section \ref{MD}). The resulting reference SOAP matrix considered for the fcc system has a size of 37044~×~50. Similarly, the reference SOAP matrix for the hcp material was built from a pure $38\times31\times19$ Zr simulation box equilibrated at 100~K, yielding a matrix of size 44764~×~105. For both materials, 10\% of the reference matrices was reserved for validation, while the remaining 90\% was split into 80\% for training and 20\% for testing using the data-splitting utilities from scikit-learn v1.3.2 \cite{scikit-learn}. Separate autoencoders were trained for the fcc and hcp materials using the PyTorch v2.2.2 framework \cite{NEURIPS2019_bdbca288} and training hyperparameters adopted for both after optimization with Optuna \cite{akibaOptunaNextgenerationHyperparameter2019} using the Tree-structured Parzen Estimator sampler on the fcc dataset. In this way, a consistent architecture was applied across materials, with the input layer size set to match the dimensionality of the SOAP vectors (i.e., 50 for fcc and 105 for hcp). Each autoencoder consists of a symmetric fully connected architecture with two hidden layers in both the encoder and decoder. The first hidden layer was set to approximately 77\% of the input dimension, the second to half the size of the first and the latent space dimension was chosen as one-fifth of the input dimension. The models were trained for 80 epochs using the Adam optimizer \cite{kingmaAdamMethodStochastic2014}, with a batch size of 256 and a learning rate of $2.74\times10^{-4}$. The mean squared error (MSE), hereafter referred to as the reconstruction error, was used as the loss function to guide the optimization process. This configuration yielded a minimum validation loss of approximately $4.3\times10^{-2}$ for the fcc autoencoder and $3.6\times10^{-3}$ for the hcp autoencoder, with no signs of overfitting observed during training.

\begin{figure}
	\centering
    \includegraphics[width=0.5\textwidth]{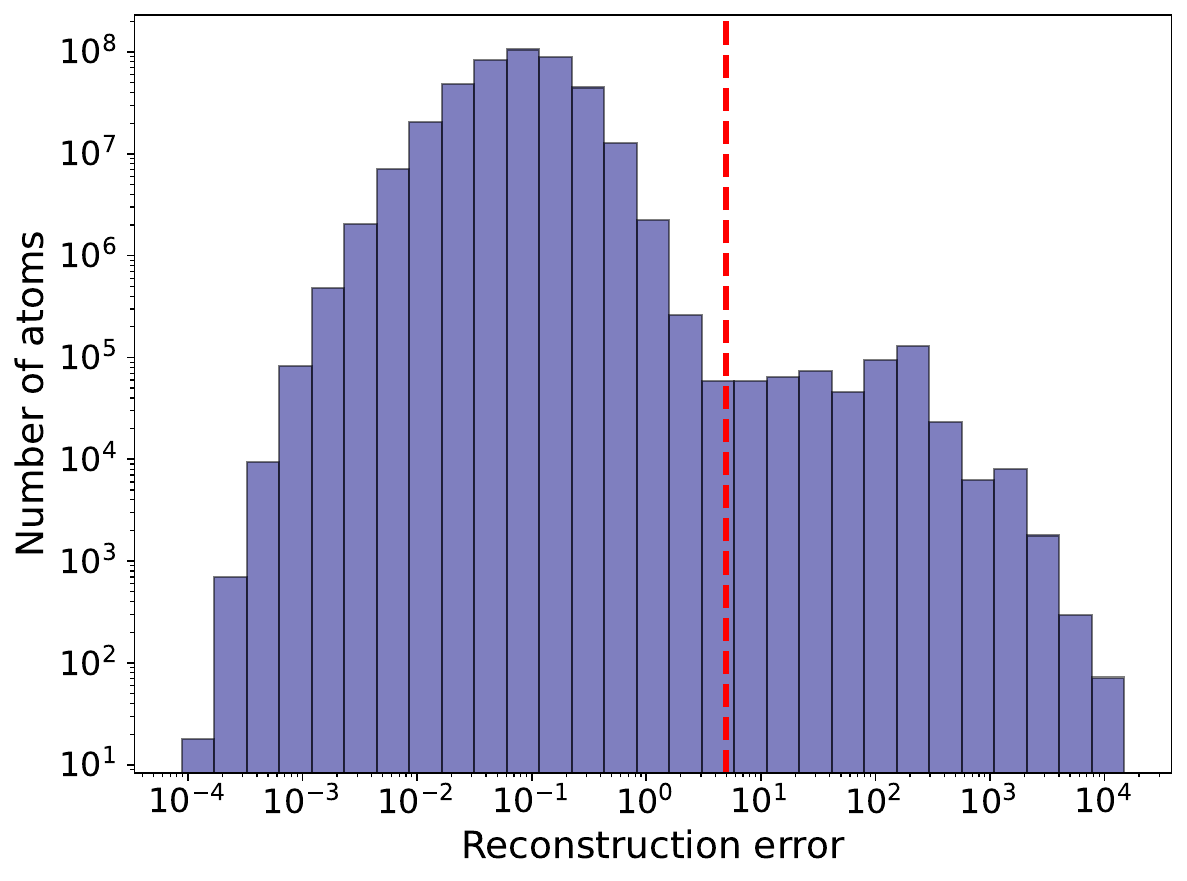}
	\caption{Log–log histogram of per-atom reconstruction errors for the Ni dataset, obtained from the autoencoder-based analysis across multiple cascade simulations. The histogram is computed using logarithmically spaced bins, and the vertical axis is also plotted on a logarithmic scale to highlight variations across several orders of magnitude. The dashed vertical line indicates the selected threshold (5.0 in this case), which lies near an inflection in the distribution, separating low-error atoms from a high-error tail linked to defective environments.}
	\label{Fig0_errordist}
\end{figure}

Once trained, the autoencoders were applied to SOAP matrices representing atomic environments at the end of the displacement cascades. Outliers were identified by evaluating the reconstruction error associated with each SOAP vector on a per-atom basis (B.2). 
Threshold values were chosen (B.3) by analyzing the log–log distribution of reconstruction errors, which exhibits a change in slope marking a qualitative shift from a broad population of low-error atoms (representative of the undisturbed crystal) to a heavy-tailed regime associated with structural anomalies or defects (see, for example, Figure~\ref{Fig0_errordist} for Ni). Guided by this observation and complemented by heuristic inspection of cascades (B.3), we selected threshold values of 5.0 for fcc and 2.0 for hcp. Atoms with reconstruction errors below the respective threshold were classified as inliers (not defective), while those above were identified as outliers and thus potentially associated with physical defects (B.4).
After outlier detection in the cascade dataset, the $N$ SOAP vectors corresponding to outlier atoms were aggregated into a single matrix for dimensionality reduction using UMAP python implementation \cite{mcinnesUMAPUniformManifold2018} (B.5). The decomposition was applied to the original $N\times50$ (fcc) or $N\times105$ (hcp) matrices, reducing them to $N\times10$ by retaining the first 10 UMAP components. The number of  approximate nearest neighbours was set to 15, while the minimum euclidean distance between embedded points was set to 0.0 to enhance group separation. Subsequently, HDBSCAN clustering was applied to the \(N \times 10\) UMAP-reduced representation of the outlier data (B.6). All parameters were kept at their default values as implemented in scikit-learn \cite{scikit-learn}, except for the minimum group size, which was set to 1000 in order to suppress the detection of what we consider statistically unrepresentative groups, i.e., those containing fewer than 1\% of the total outlier population.

\subsection{Conventional characterization of point defect and clusters}

To provide a statistical comparison with the machine learning approach, several conventional defect identification methods were considered.
The classical lattice-site analysis consists in evaluating atom occupancy around each ideal lattice position using an enhanced in–out touching-sphere criterion \cite{debackerModellingPrimaryDamage2021, nairStatisticalStudyDisplacement2025, debackerModelDefectCluster2018}: atoms are counted within a central touching sphere and a narrow surrounding shell, allowing the identification of vacancies and self-interstitials located just beyond their nominal lattice cages. As this method relies on a perfect reference lattice, its accuracy may deteriorate in regions exhibiting significant structural distortion. 
In addition, we also applied the standard local-environment algorithms available in \textsc{OVITO} \cite{ovito}: centrosymmetry parameter (CS), coordination analysis (CA), polyhedral template matching (PTM), dislocation extraction algorithm (DXA) and Voronoi analysis (VOR). All the parameters are described hereafter in table~\ref{tab:method_summary}.

\color{black}

\section{Results and discussion}\label{full_results}

The datasets analyzed in this study consist of around 110 displacement cascades per material, each initiated with a PKA energy of 80~keV. The FeNiCr alloy investigated has a composition of 70~at.\% Fe, 10~at.\% Ni, and 20~at.\% Cr.

\subsection{Defects detection using the ML pipeline}\label{ML_res_detect}

\subsubsection{Outliers identification with autoencoders}

\begin{figure}
	\centering
    \includegraphics[width=\textwidth]{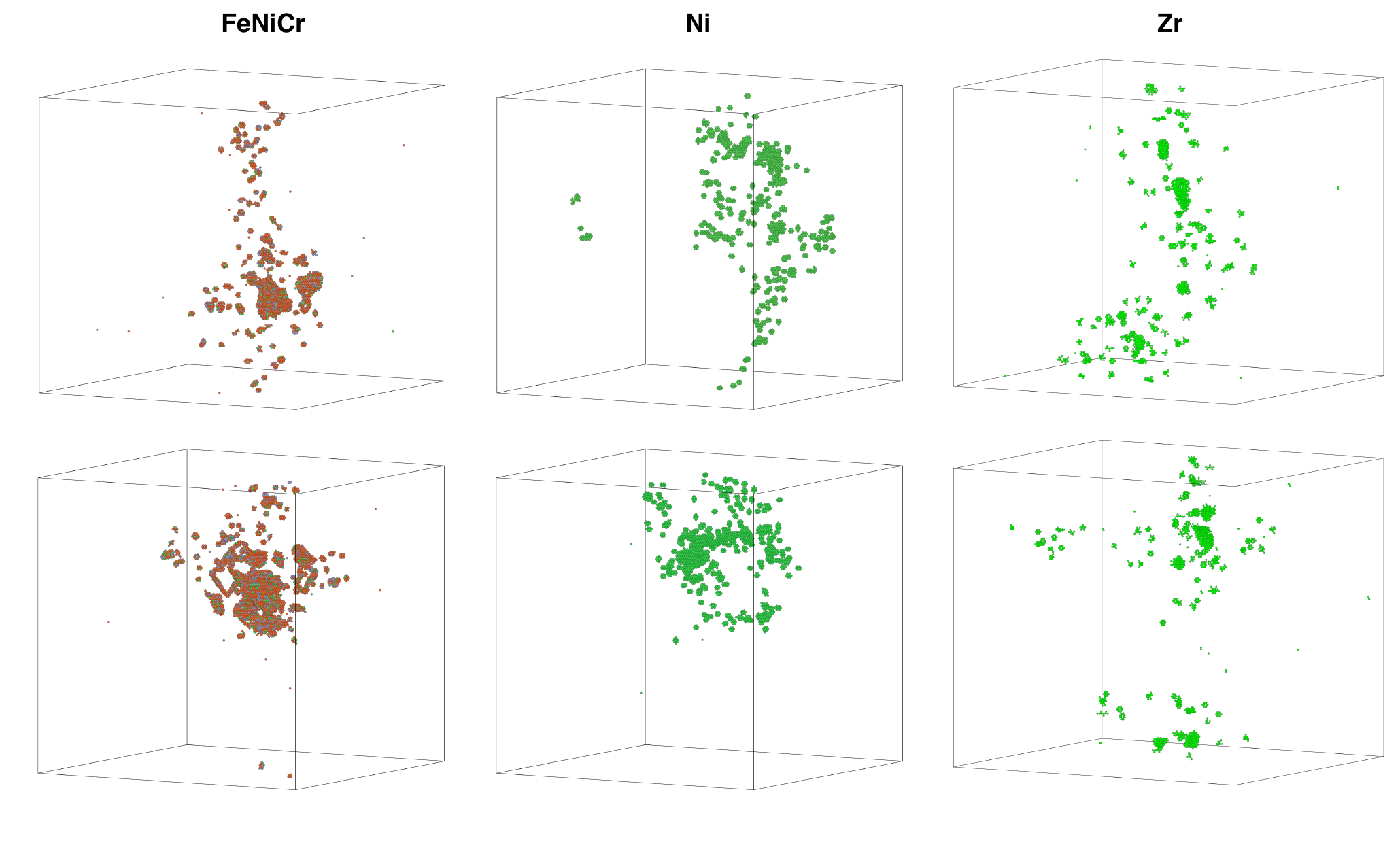}
	\caption{Example of outlier detection in FeNiCr (left), Ni (middle) and Zr (right) displacement cascades (step B.4, see Figure \ref{Fig0_diag}). Atoms whose auto-encoder reconstruction error is below 5.0 (fcc systems) or 2.0 (hcp system) are omitted from the view; only atoms flagged as outliers are shown in solid colors.}
	\label{Fig1}
\end{figure}

Figure \ref{Fig1} presents an illustrative example of outlier identification performed using the autoencoders in typical displacement cascade simulations for FeNiCr, Ni, and Zr (step B.4). It should be noted that cascade properties and morphologies are subject to fluctuations, as reported for instance in \cite{nairStatisticalStudyDisplacement2025} for fcc materials, and Figure \ref{Fig1} therefore illustrates only a single representative example for each material.
In each case, atoms classified as normal (i.e., those whose local environments closely match the reference crystalline structure) are not rendered and atoms identified as outliers based on their reconstruction error are displayed in solid colors, highlighting their deviation from the bulk structure. 
This first step of the machine learning workflow demonstrates that the autoencoders can accurately distinguish between defect-free and perturbed local atomic environments in both fcc and hcp structures, even though defective atoms represent only a small fraction of the total simulation box. This highlights the model's sensitivity to subtle local distortions and its robustness in detecting structural anomalies within a predominantly crystalline background. In particular, a first visual inspection of the atomic environments reveals that both the peripheral atoms surrounding defective sites and those located at the core of defective environments are highlighted, both in isolated defects and within clustered regions. Importantly, this detection is achieved without any prior knowledge of defect types or geometries, making the approach fully unsupervised and adaptable to unknown or complex defect morphologies. An initial observation can be made regarding the fcc materials, where FeNiCr cascades tend to exhibit more large and aggregated defect structures compared to pure Ni. On the other hand, Zr cascades appear to generate fewer defects overall, with most of them being spatially isolated. 

\subsubsection{Unsupervised mapping of defect morphologies and visual inspection of HDBSCAN groups}

\begin{figure}
	\centering
    \includegraphics[width=\textwidth]{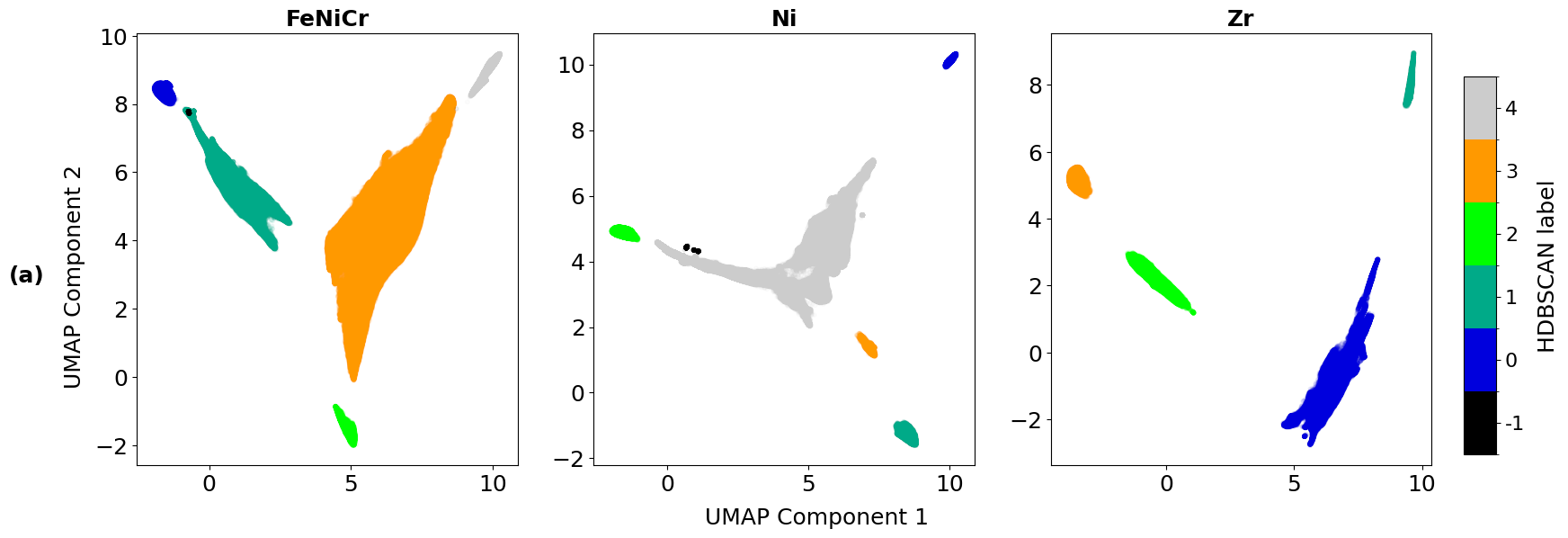}
    \includegraphics[width=0.9\textwidth]{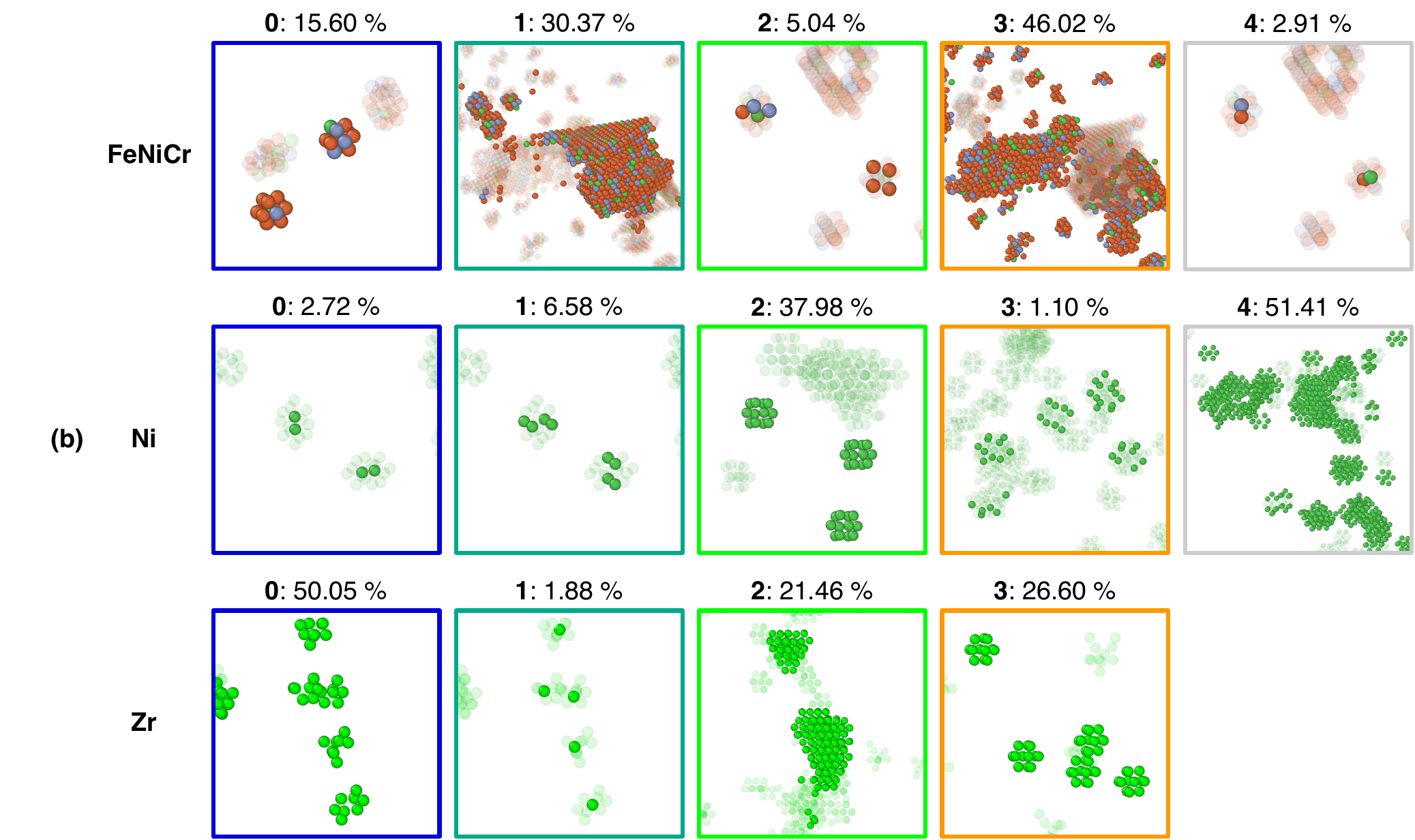}
	\caption{(a) Two-dimensional UMAP projection of latent-space SOAP descriptors for outliers in FeNiCr, Ni and Zr systems, colored by HDBSCAN group labels. For FeNiCr and Ni, points labeled as -1 (black) represent samples that HDBSCAN did not assign to any group. Note that the HDBSCAN labels are assigned independently for each material, meaning that the same label in different systems does not necessarily correspond to the same type of defect pattern (b) Examples of representative atomic configurations associated with selected HDBSCAN groups from (a) (label -1 excluded), shown using the same color scheme. The relative size of each group, as a percentage of the outlier dataset, is also indicated. Transparent atoms represent outlier atoms associated to other HDBSCAN labels.}
	\label{Fig2}
\end{figure}

To further investigate defect patterns and allow their unsupervised classification at larger scale, Figure \ref{Fig2} (a) presents the projection of the SOAP descriptors corresponding to outlier atoms in the FeNiCr, Ni, and Zr cascade datasets onto the first two UMAP components (step B.5). Points are colored according to the HDBSCAN group labels assigned independently for each material, revealing structurally similar groups of atomic environments in an unsupervised manner (step B.6). Figure \ref{Fig2}(b) displays representative atomic configurations corresponding to the HDBSCAN groups identified in Figure \ref{Fig2}(a) for each material. 
For all three materials, the combination of UMAP projection and HDBSCAN clustering partitions the outlier population into compact, well-defined groups in the UMAP latent space offering a quantitative map of the structural diversity captured in the outlier population. Despite the fully unsupervised nature of the workflow, fewer than 0.05\% of the FeNiCr outliers and 0.22\% of the Ni outliers remain unassigned (HDBSCAN label –1), and all Zr outliers are grouped. This confirms that almost every anomalous atom can be associated with a recurring local environment motif. It should be noted that, since the machine learning workflow is entirely unsupervised, the nature of each group cannot be inferred directly from the algorithm output; visual inspection of the atomic environments is therefore a necessary first step. This qualitative analysis serves as a basis for the more quantitative comparison with conventional defect identification methods presented in the following section.

Group size statistics provide further insight into the structural landscape revealed by the ML workflow.
In FeNiCr, five HDBSCAN groups (labels 0–4) are identified. Labels 3 and 1 dominate, accounting for 46.0~\% and 30.4~\% of all outlier atoms, respectively, whereas labels 0 (15.6~\%), 2 (5.0~\%), and 4 (2.9~\%) represent progressively smaller fractions of the outlier population. In addition, none of the HDBSCAN‑defined groups exhibit statistically significant enrichment in Fe, Ni or Cr.
Representative snapshots extracted from Figure \ref{Fig2}(b) clarify the structural meaning of these groups. Groups 0 consists almost entirely of the twelve first-neighbour atoms that form the cage around a monovacancy. Group 4 contains the two atoms that make up a dumbbell, while group 2 comprises the nearest-neighbour atoms located perpendicular to the dumbbell axis and slightly displaced by its presence.
The two most populated categories, groups 3 and 1, correspond predominantly to large, aggregated defect structures whose precise character (interstitial-rich versus vacancy-rich) can not be determined only from atom positions but complementary evidence offers useful clues. In the UMAP embedding, group 3 lies between groups 4 and 2, both linked to dumbbell interstitials and their first neighbours. In the atomistic snapshots, group 3 atoms appear not only within extended defect regions but also as a second-shell “halo” surrounding individual dumbbells. Taken together, these observations suggest that group 3 is dominated by interstitial-type aggregates, ranging from isolated dumbbells to larger interstitial complexes.
On the other hand, visual inspection of atom configurations from group 1 reveals mainly large, multi-atom structures, and its position adjacent to group 0 in UMAP space further suggests that group 1 is vacancy-rich and most likely captures aggregated vacancy complexes. A definitive classification of both groups (3 and 1), however, will rely on the spatial metrics and defect counting presented in Section \ref{spatial_clust}.

Regarding Ni, five HDBSCAN groups (labels 0–4) are also obtained, but their size distribution differs significantly from that of FeNiCr. Two groups dominate: label 4 contains 51.4 \% of all outlier atoms, and label 2 accounts for a further 38.0 \%. The remaining groups labeled 1 (6.6 \%), 0 (2.7 \%), and 3 (1.1 \%) together contribute less than 11 \% of the outlier population. A clear correspondence emerges between the two fcc systems: in Ni, HDBSCAN groups 2, 0, and 1 align with groups 0, 4, and 2 in FeNiCr, respectively. These groups capture \emph{(i)} the twelve nearest-neighbour atoms surrounding a vacancy or vacancy-like cavity, \emph{(ii)} the two atoms forming a dumbbell interstitial, and \emph{(iii)} the first-neighbour atoms lying in the plane perpendicular to the dumbbell axis. This correspondence is visible in the UMAP projections: in both fcc datasets, the groups associated with monovacancies and vacancy-centred environments (Ni : group 2; FeNiCr : group 0) fall in the upper-left region of the map. Likewise, the dumbbell interstitials (Ni : group 0; FeNiCr : group 4) occupy the upper-right quadrant, and the first-neighbour atoms perturbed by the dumbbells (Ni : group 1; FeNiCr : group 2) are concentrated toward the lower portion of the embedding. This spatial alignment confirms that UMAP preserves local similarity well enough for crystallographically equivalent defect motifs from separate datasets to appear in roughly the same regions of the UMAP latent plane. Group 4 displays the same dual signature observed for group 3 in FeNiCr: it includes both the second-shell “halo” atoms capping dumbbells and atoms that belong to larger, aggregated defects likely attributed to interstitials type complexes. However, the broad horizontal spread of group 4 across the UMAP plane (extending from the vacancy-rich area (adjacent to group 2) to the interstitial-dominated region (near groups 0, 1, and 3)) suggests that, compared with group 3 of FeNiCr, group 4 in Ni is more likely to present a mixture of interstitial- and vacancy-type aggregates rather than a purely interstitial motif.

Interestingly, most atoms assigned to group 3 form organized shells surrounding what appears to be icosahedral-type structures. Moreover, the position of group 3 in the UMAP projection located in the bottom-right region between groups 1 and 4 suggests that it likely corresponds to an interstitial-rich defect environment. This group may therefore be associated with the presence of A15-type structural motifs \cite{frankComplexAlloyStructures1959,goryaevaCompactA15FrankKasper2023}. The correlation between this group and the results of the PTM analysis which can highlight icosahedral environment associated to the presence of A15 phases \cite{wyszkowskaNanoscaleDefectFormation2025}, will be discussed hereafter.

Finally, four HDBSCAN groups are highlighted for Zr. Group 0 dominates the outlier population, including approximately half of all anomalous atoms (50.1~\%). Two additional groups, 2 and 3, account for 21.5~\% and 26.6~\%, respectively, while the smallest group, labelled 1, represents only 1.9~\%. This later group according to Figure \ref{Fig2}(b) corresponds predominantly to single atom that is therefore interpreted as a population of single interstitial defects. Group 0 is then best interpreted as the first-neighbour atoms surrounding the interstitials. This assignment is consistent with the UMAP embedding, where group 0 occupies the same sector of the map as group 1, already attributed to single interstitials. As in the fcc cases (Ni : group 2; FeNiCr : group 0), group 3 in Zr is primarily associated with monovacancies, appearing as the cage of first-neighbour atoms that surrounds an empty lattice site. Group 2 shows more extended defect assemblies and its position adjacent to group 1 in the UMAP map suggests that it is also vacancy-rich.

\subsection{Comparison with conventional methods}

\subsubsection{Local spatial cluster analysis on ML-detected outliers}\label{spatial_clust}

\begin{figure}
	\centering
    \includegraphics[width=\textwidth]{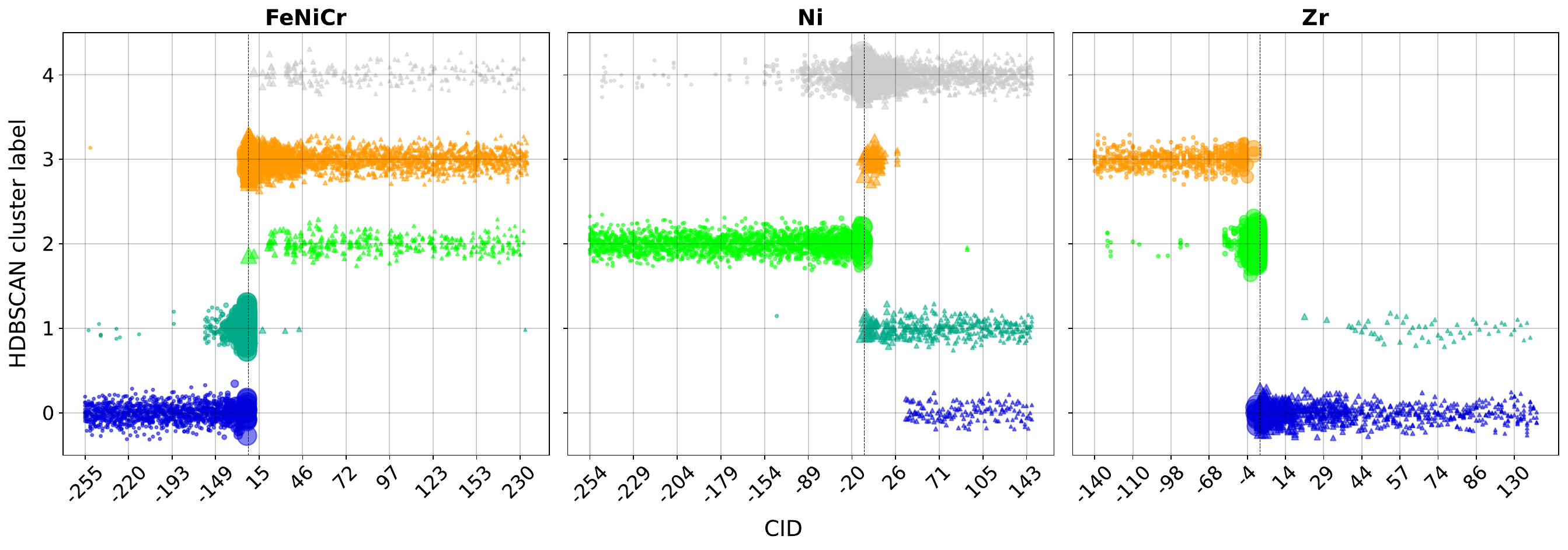}
    \caption{Cluster-identification (CID, see section \ref{spatial_clust} for description) diagnostics for the HDBSCAN groups for one typical cascade for FeNiCr, Ni and Zr. For each material, the histogram displays the distribution of the variable \(\mathrm{CID}=\mathrm{sign}(n_{\text{Def}})\times\mathrm{DefID}\), where \(\mathrm{sign}(n_{\text{Def}})>0\) (triangles) denotes interstitial-dominated clusters and \(\mathrm{sign}(n_{\text{Def}})<0\) (circles) denotes vacancy-dominated clusters.  The vertical dotted line marks \(\mathrm{CID}=0\).  The magnitude \(|\mathrm{CID}|\) is inversely proportional to aggregate size: small \(|\mathrm{CID}|\) corresponds to large clusters, large \(|\mathrm{CID}|\) to small clusters and marker size is inversely proportional to \(|\mathrm{CID}|\). Hence, points far to the right represent small interstitial defects, whereas points far to the left correspond to small vacancy defects; values near the origin indicate the largest aggregates of either type. Bars are coloured according to the HDBSCAN labels defined in Fig.~\ref{Fig2}(a).}
	\label{Fig3}
\end{figure}

To refine the physical interpretation of each HDBSCAN group, we applied a spatial–clustering analysis to all atoms flagged as outliers for each material, using a cut-off radius of \(r_{\mathrm{c}} = 3.6\)~Å and $3.5$~Å for fcc and hcp materials, respectively.  Each spatial cluster is assigned an identification number by decreasing size order (DefID): low DefID values correspond to large aggregates, while high values correspond to small ones. For every cluster we record both its total atom count and its point-defect composition. Vacancies and interstitials are classified within each spatial cluster with the aforementioned classical lattice site analysis, counting interstitials as \(+1\) and vacancies as \(-1\); the algebraic sum yields a defect score ($n_{\mathrm{Def}}$) whose sign indicates whether the cluster is vacancy-dominated (negative) or interstitial-dominated (positive).  Multiplying this sign by the DefID produces a single diagnostic quantity hereafter called the cluster identification (CID) in which the sign discriminates interstitial versus vacancy character, while the magnitude is inversely proportional to aggregate size (small absolute values for large clusters, large absolute values for small clusters). Figure \ref{Fig3} plots the distribution of this diagnostic for every HDBSCAN label for one typical representative cascade. It should be noted that similar characteristics are observed across the other cascades for each material.
Groups 4 and 2 in FeNiCr show predominantly large positive CIDs, confirming that they are composed of small-to-medium size interstitial clusters that is consistent with the dumbbell interstitials themselves (groups 4, high CID) and their first-neighbour “halo” atoms (group 2, moderate CID) highlighted using the ML approach. Group 3, which dominated the outlier population in Fig. \ref{Fig2}~(a), also forms band of mainly low-magnitude positive CIDs, which is consistent with interstitial-rich, large aggregates. Conversely, groups 0 and 1 peak on the negative side: group 0 at small-to-moderate CID corresponds to small-to-medium vacancies, whereas group 1 mostly spans over small CID values, identifying the vacancy-dominated counterpart of group 3 (large vacancy aggregates). The clean separation of vacancy- and interstitial-dominated signatures substantiates the UMAP-based assignment proposed earlier.
For Ni, group 0 appears at large positive CID, matching the core atoms of dumbbell interstitials; group 1 lies at moderate positive values, as expected for moderate-size interstitials complexes consistent with the neighbouring atoms displaced by those dumbbells.
Group 2 occupies the most negative CID range, confirming ML prediction that it is a vacancy-dominated aggregate of all sizes. Group 4 spreads across the origin, displaying both positive and negative tails with a skew toward positive CID values. This asymmetry implies that although the group contains some vacancy-type clusters (negative tail), it is predominantly interstitial-rich. 
The mixed signature corroborates the UMAP observation that group 4 lies between the vacancy and interstitial regions yet leans closer to the interstitial side, distinguishing it from the purely interstitial analogue seen in FeNiCr. Group 3 appears to be well located on the low-magnitude positive CID values. The narrow spread of these low positive values further suggests a fairly uniform size, consistent with the concentric “shell-in-shell” arrangement observed in the snapshots and characteristic of A15-type interstitial complexes.
In the hcp metal Zr the CID histogram resolves four well-defined families. Group 1 lies at the positive end, consistent with isolated interstitials. Group 0 follows mainly at small and moderate positive CID values that reveals the signature of both large and moderate interstitial type complexes, in line with Fig. \ref{Fig2} (b) which mainly represents the first-neighbour atoms around those interstitials.
On the vacancy side, group 3 peaks at large to moderate negative magnitudes corresponding to small and moderate size vacancy complexes, coherent with ML observation of predominantly monovacancies, while group 2 is centered and localised around low-magnitude negative CID, indicative of extended vacancy aggregates. The narrow spread suggests relatively similar aggregate sizes. The absence of overlap between positive and negative regions shows that the AE → UMAP → HDBSCAN pipeline cleanly distinguishes interstitial- and vacancy-type defects even in the anisotropic hcp lattice.

Taken together, the CID histograms corroborate the earlier UMAP-based assignments: each HDBSCAN group forms a coherent physical motif whose vacancy or interstitial character and approximate size are now confirmed quantitatively. 
Particularly noteworthy is the asymmetric spread of group 4 for Ni, which shows that this group contains a substantial interstitial component but also a detectable vacancy fraction. This insight could not be highlighted from the UMAP projection and structure visualisation alone, however, the use of CID histograms such as Fig. \ref{Fig3} can provide valuable additional information that can complete the ML framework.

\begin{figure}
	\centering
    \includegraphics[width=\textwidth]{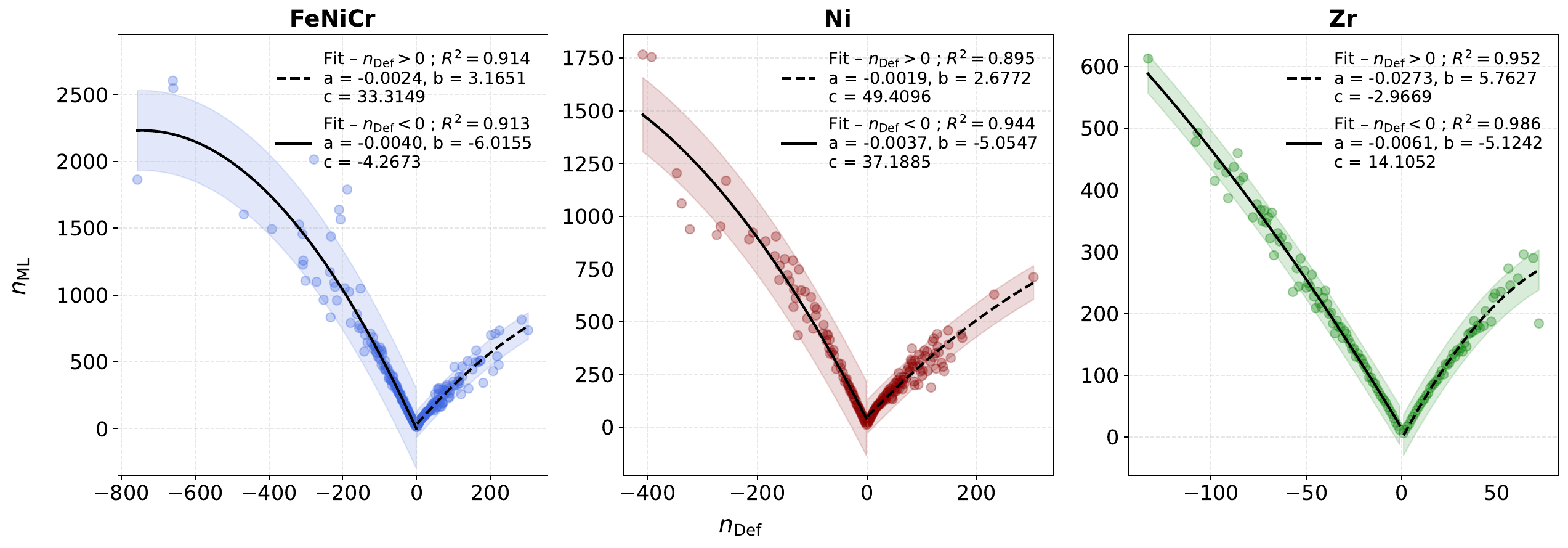}
	\caption{Correlation between the number of outlier atoms per cluster ($n_{\mathrm{ML}}$, as identified by the machine learning approach) and the number of defects ($n_{\mathrm{Def}}$) within each cluster for the three materials, with $n_{\mathrm{Def}}$ defined as the signed sum of defects (interstitial counted as $+1$ and vacancy as $-1$). Second-order polynomial fits are used to calibrate the relationship, enabling estimation of either the number of atoms or the number of defects in a cluster.  Fits are performed separately for interstitial-rich ($n_{\mathrm{Def}} > 0$, dashed lines) and vacancy-rich ($n_{\mathrm{Def}} < 0$, solid lines) clusters.  Shaded regions indicate $\pm 2\sigma$ confidence intervals. Fitted equations and corresponding $R^2$ values are shown in the legend, where $a$ quantifies the quadratic contribution, $b$ the linear contribution, and $c$ the constant offset.}
	\label{Fig4}
\end{figure}

Figure \ref{Fig4} reveals a smooth relationship between the average number of outlier atoms per cluster $(n_{\mathrm{ML}}$ detected with the ML workflow, and the net defect content $n_{\mathrm{Def}}$ of that cluster computed as described above.  For each material, the correlations are captured with good accuracy ($R^{2}>0.89$) by two independent second-order polynomials, fitted separately on the vacancy-rich branch $(n_{\mathrm{Def}}<0)$ and on the interstitial-rich branch $(n_{\mathrm{Def}}>0)$.  These fits constitute practical calibration curves since knowing $n_{\mathrm{ML}}$ one can estimate how many defects belong to a given cluster, and a known defect inventory can be translated into the number of atoms constituting a given aggregate.
In addition, the polynomial coefficients encode useful physics.  In the low–defect limit where the quadratic term \(a\) is negligible, the the linear coefficient \(b\) ($b_{\text{int}}$ for $n_{\mathrm{Def}}>0$, and $b_{\text{vac}}$ for $n_{\mathrm{Def}}<0$) provides a direct approximation of the number of atoms $n_{\mathrm{ML}}$ forming a defect aggregate as a function of its net defect content $n_{\mathrm{Def}}$. In this case, we adopt a single calibration law for each defect type (one for interstitials and one for vacancies), rather than distinguishing multiple regimes that may correspond to different defect families. Inspection of the plots, however, indicates that such regime separation does not seem to play a significant role here. A more detailed discrimination between different categories of defect clusters (e.g., pyramidal vs. loop types) will be addressed in forthcoming studies on Zr \cite{Debacker_zr2025}.
In FeNiCr alloy we obtain $b_{\text{int}}\approx 3.17$ and $|b_{\text{vac}}|\approx 6.02$, so a vacancy perturbs roughly two times more atoms than an interstitial. Pure~Ni shows the same trend but with a smaller ratio, $b_{\text{int}}/|b_{\text{vac}}|\approx 0.53$, indicating that vacancies remain the most disruptive defect but the
pure Ni lattice accommodates their strain slightly more efficiently than the alloy (as evidenced by its smaller interstitial slope, $b_{\text{int}}\!\approx\!2.68$, and less-negative vacancy intercept, $c_{\text{vac}}\!\approx\!-37$). This is in line with expectations for a monatomic, nearly isotropic fcc lattice in which all sites share the same elastic properties and atomic size. 
The hcp metal Zr exhibits $b_{\text{int}}\!\approx\!5.76$ and $|b_{\text{vac}}|\!\approx\!5.12$, that gives a ratio of $\approx 1.12$.  Thus, in Zr an interstitial disturbs nearly as many atoms as a vacancy.
Finally, the negative quadratic term \(a\) accounts for the curve’s flattening at large \(|n_{\mathrm{Def}}|\).  In densely defected clusters the local distortions overlap, so each additional defect perturbs fewer new atoms and \(n_{\mathrm{ML}}\) increases more slowly. Geometrically, the number of disturbed sites scales with the interface area of the aggregate rather than with its volume, so the growth of \(n_{\mathrm{ML}}\) becomes sublinear. 
Overall, on the vacancy‑rich side the largest aggregate in the FeNiCr material contains $\sim2600$ outlier atoms at $n_{\mathrm{Def}}\!\approx\!-780$.  Pure Ni tops out around $\sim1700$ outliers for $n_{\mathrm{Def}}\!\approx\!-390$, whereas Zr levels off near $\sim600$ outliers for $n_{\mathrm{Def}}\!\approx\!-120$.  The same hierarchy holds on the interstitial branch: FeNiCr rises to 950 outliers at $n_{\mathrm{Def}}\!\approx\!+230$, Ni to 700 at $+260$, and Zr to 300 at $+60$.  Hence the two fcc materials produce defect aggregates that are roughly three to four times larger than those in the hcp metal, and the fcc alloy outgrows its pure‑Ni counterpart.  This numerical ranking of aggregate sizes mirrors the extended vacancy agglomerates visible in Fig.~\ref{Fig1} and \ref{Fig2}.

\subsubsection{Number of detected outliers}

\begin{table}[htbp]
\centering
\caption{Inputs, per‑atom outputs and “normal–atom’’ removal (inliers) rules for every detector used in this study. VI = Voronoi index. Note that preliminary tests were performed to support these choices, while a full sensitivity analysis lies beyond the scope of this paper.}
\label{tab:method_summary}
\renewcommand{\arraystretch}{1.15}
\begin{tabular}{|p{1.65cm}|p{3.95cm}|p{4.00cm}|p{4.05cm}|}
\hline
\textbf{Method} &
\textbf{Key input / parameters} &
\textbf{Per‑atom output} &
\textbf{Normal‑atom filter} \\
\hline\hline
CS &
12 nearest neighbours (fcc and hcp) &
Float: centrosymmetry parameter &
Discard atoms with CS~<~1.8 for fcc and |CS-9.6|~<~0.2 for hcp \\
\hline
CA &
Cut‑off radius $R_\text{c}=4.7$~Å (fcc and hcp) &
Integer coordination &
Discard atoms with $coordination = 42$ for fcc and $coordination = 18$ for hcp \\
\hline
VOR &
\begin{itemize}[nosep,label*={},leftmargin=*]
    \item Atomic volume
    \item nearest‑neighbour list
    \item VI template set
\end{itemize} &
\begin{itemize}[nosep,label*={},leftmargin=*]
\item Atomic volume
\item complete VI
\item max face order
\end{itemize}&
Discard atoms of atomic volume $(a^{3}/4) < 0.7$\AA$^3$ for fcc and atoms with the Voronoi polyhedron of the perfect crystal for Zr\\
\hline
PTM &
\begin{itemize}[nosep,label*={},leftmargin=*]
    \item Template library
    \item RMSD tolerance (0.60~Å)
\end{itemize}
&
\begin{itemize}[nosep,label*={},leftmargin=*]
\item Template label (fcc, hcp, bcc, …)
\item RMSD
\end{itemize}
&
Discard atoms assigned to fcc or hcp template \\
\hline
DXA &
\begin{itemize}[nosep,label*={},leftmargin=*]
\item Trial circuit length 
\item circuit stretchability 
\end{itemize}
&
\begin{itemize}[nosep,label*={},leftmargin=*]
\item Dislocation segments 
\item local crystal type 
\end{itemize}&
Discard atoms classified as perfect fcc or hcp \\
\hline
\end{tabular}
\end{table}

Table~\ref{tab:method_summary} shows how each conventional scheme can be turned into an outlier detector. For every method it indicates \emph{(i)} the minimal input it needs, \emph{(ii)} the per‑atom quantity it computes and \emph{(iii)} the numerical cut‑off or template match that classifies an atom as “normal” (inlier) and therefore excluded. By juxtaposing the different filters, the table shows how one converts raw structural descriptors into a binary decision and highlights what is required to detect outliers in lattices beyond fcc or hcp. 

\begin{table}[ht]\label{outlier_rates_all_methods}
\centering
\caption{Outlier fractions (\% of total number of atoms in each dataset) reported by the ML workflow and each classical reference method, together with the total dataset size for each alloy.}
\label{tab:outlierRates}
\begin{tabular}{lrrrrrrrr}
\toprule
Material & Atoms per dataset & ML & CS & CA & DXA & PTM & VOR \\
\midrule
FeNiCr  & 417,889,248 & 0.1271 & 0.1637 & 0.3247 & 0.1507 & 0.0772 & 0.0766 \\
Ni      & 417,889,248 & 0.1236 & 0.1356 & 0.3324 & 0.1343 & 0.0789 & 0.0627  \\
Zr      & 307,745,280 & 0.0979 & 0.0596 & 0.5385 & 0.1211 & 0.0363 & 0.6713  \\
\bottomrule
\end{tabular}
\end{table}

Table \ref{tab:outlierRates} reports, for each system, the percentage of the entire atom population that each detector classifies as an outlier, alongside the total number of atoms in the datasets. Although the absolute outlier fractions are threshold and parameters dependent, they still reveal some physically interpretable trends. First, all methods agree that surviving damage is extremely sparse: even the most liberal detector (VOR in Zr) flags less than 0.7~\% of sites, while the most conservative (PTM in Zr) drops below 0.04~\%. Such sub-percent residual-defect densities are entirely consistent with molecular dynamics cascade studies at similar recoil energies, substantiating that the thresholds have not driven any method into an obviously unphysical regime. Second, the relative ordering of methods is systematic across alloys. PTM is always the most stringent followed closely by the Voronoi indices in the fcc systems; CS and DXA cluster tightly around the ML baseline; CA consistently inflates the count by a factor $\approx$2.5 in fcc alloys and by more than 5 in the hcp alloy. Because these rank positions persist in all three datasets, they reflect systematic differences in algorithmic sensitivity.
Third, crystal structure modulates the alarm volume in opposite ways for the two detector families. Neighbour-geometry–based algorithms (PTM, CS, DXA) and ML approach  register fewer outliers in hcp Zr than in fcc Ni or FeNiCr, which is consistent with observations described above (see section \ref{full_results}). By contrast, CA and the Voronoi metrics report far more outliers in Zr (up to seven-fold). 
Finally, the two fcc datasets differ by no more than 0.03 percentage-points for any method, implying that alloying introduces only a marginal increase in static lattice distortion at the simulated irradiation energy.
In sum, while one should avoid comparing absolute numbers across detectors with dissimilar thresholds, the relative ordering and the crystal-structure dependence visible in Table \ref{tab:outlierRates} are indicators of each algorithm’s intrinsic sensitivity and of the underlying damage landscape. A more detailed comparative statistical analysis follows in the next section. 

\subsubsection{Statistical relation between outliers detection using conventional methods and ML}

\begin{table}[ht]
\centering
\caption{Global precision (P), recall (R) and $F_1$-score of the ML workflow relative to six conventional reference methods in the three systems}
\label{tab:PRF}
\begin{tabular}{lccc}
\toprule
Method & FeNiCr: P / R / $F_1$ & Ni: P / R / $F_1$ & Zr: P / R / $F_1$ \\
\midrule
CS    & 0.935 / 0.727 / 0.816 & 0.968 / 0.882 / 0.923 & 0.426 / 0.698 / 0.528 \\
CA    & 0.840 / 0.329 / 0.474 & 0.876 / 0.326 / 0.475 & 0.203 / 0.130 / 0.159 \\
DXA   & 0.954 / 0.805 / 0.873 & 0.980 / 0.902 / 0.939 & 0.884 / 0.718 / 0.792 \\
PTM   & 0.573 / 0.944 / 0.713 & 0.621 / 0.974 / 0.759 & 0.338 / 0.917 / 0.494 \\
VOR   & 0.416 / 0.690 / 0.519 & 0.445 / 0.876 / 0.590 & 0.023 / 0.887 / 0.045 \\
\bottomrule
\end{tabular}
\end{table}

For a first global statistical view, Table \ref{tab:PRF} reports the global precision (P), global recall (R), and $F_1$-score of the ML workflow with respect to the five conventional reference methods (CS, CA, DXA, PTM, and VOR) across the three simulated systems. These performance metrics are derived from the standard confusion-matrix framework. In this context,
true positives (TP) are atoms that both the conventional method and ML flag as anomalies. False positives (FP) are atoms the ML model declares outliers even though the reference method judges them normal. False negatives (FN) are atoms the reference method tags as anomalous while ML misses them. The remaining atoms, classified as normal by both approaches, form the true negatives (TN) and constitute the overwhelming majority because only around 0.1~\% of the cascade-simulation atoms are defective.
Then, global recall  (true-positive rate, R = TP/(TP+FN)) reports the fraction of reference-defined outliers that the ML workflow actually recovers. Global precision (positive-predictive value, P = TP/(TP+FP)) indicates how often an atom flagged as outlier by ML is confirmed by the conventional method. 
In addition, their harmonic mean is the $F_1$ score ($F_1 = 2PR/(P+R)$), which balances recall (completeness) and precision (reliability). This is particularly useful under the extreme class imbalance of our problem, where only about one atom in a thousand is flagged as an outlier.
The primary objective of the matrix analysis is therefore to determine how faithfully the ML detector reproduces each conventional definition of an “outlier” across chemically and structurally distinct systems while accounting for the around $10^{-3}$ outlier fraction characteristic of displacement-cascade datasets.
It should be noted, however, that the values reported in Table~\ref{tab:PRF} depend on specific parameter choices (see table \ref{tab:method_summary}). While a full sensitivity study lies beyond the scope of this work, the table is intended to provide an overview of the general trends and relative performance of the methods with respect to the ML approach.
In the two fcc materials (FeNiCr and Ni) the ML labels coincide most closely with those from the DXA and CS, yielding precision values of $0.94$–$0.98$ and recall values of $0.73$–$0.90$, so that more than four-fifths of all DXA/CS outliers are recovered by ML while very few additional atoms are flagged.  The correspondence with DXA remains strong in the hcp Zr cascade ($P=0.884$, $R=0.718$), whereas CS shows a marked drop in precision ($P=0.426$) because, with the chosen normal-atom filter (see table \ref{tab:outlierRates}), CS itself labels fewer Zr atoms as defective. Relative to PTM, the ML workflow exhibits very high recall ($\ge 0.92$ in all alloys) but only moderate precision ($0.34$–$0.62$), indicating that virtually every PTM-labelled defect is retained, while the ML model adds further sites that the stricter PTM template match omits.  In contrast, when benchmarked against CA, the ML detector shows low recall (0.13–0.33) and only moderate-to-high precision in the fcc systems (0.84–0.88) that drops in Zr (0.203). Relative to the Voronoi topological metrics (VOR), however, it attains respectable recall (0.69–0.88 in the fcc metals and 0.89 in Zr) yet low precision, especially in Zr (P=0.023).
This difference arises from the choice of the normal-atom filter in the Voronoi method for fcc and hcp systems (see Table \ref{tab:method_summary}). For fcc, we applied a heuristic criterion based on the atomic volume, whereas for hcp we relied on the theoretical Voronoi polyhedral shape of the crystal lattice, a filter that ultimately proved to be poorly selective. In the CA method, the normal-atom filter is based on the coordination number, which is an integer and therefore does not allow for a smooth threshold. In this case, we adopted the conservative choice corresponding to the perfect crystal. This results in much larger outlier fractions reported by CA in both fcc and hcp lattices, and by Voronoi in hcp (Table \ref{tab:outlierRates}) so ML accepts only a subset of their expansive defect catalogues. Overall, the ML scheme reproduces the moderate-volume, high-specificity behaviour of DXA and CS, extends the conservative PTM catalogue because of the large RMSD, and offers a far tighter selection than the expansive CA (fcc and hcp) and Voronoi (hcp) approaches. Within a single parameter set it therefore achieves a balance between sensitivity and specificity comparable to, or better balanced than, any individual conventional detector across both fcc and hcp crystal structures.

\begin{table}[ht]
\centering
\caption{Recall of six conventional detectors for every ML‑defined HDBSCAN label (noise group (label -1) omitted).  
Each value is the fraction of ML‑flagged atoms in the group that the detector also classifies as outliers.}
\label{tab:clusterRecallFull}
\begin{tabular}{lrrrrrrr}
\toprule
Alloy & Group & CS & CA & DXA & PTM & VOR\\
\midrule
FeNiCr & 0 & 1.000 & 0.986 & 1.000 & 0.776 & 0.686 \\
       & 1 & 0.898 & 0.972 & 0.923 & 0.700 & 0.505 \\
       & 2 & 0.998 & 0.933 & 1.000 & 0.115 & 0.716 \\
       & 3 & 0.931 & 0.683 & 0.951 & 0.444 & 0.196 \\
       & 4 & 1.000 & 0.976 & 1.000 & 0.998 & 0.986 \\
\midrule
Ni     & 0 & 1.000 & 0.995 & 1.000 & 0.999 & 1.000 \\
       & 1 & 0.999 & 0.980 & 1.000 & 0.124 & 0.736  \\
       & 2 &  1.000 & 0.973 & 1.000 & 0.804 & 0.529 \\
       & 3 & 1.000 & 0.960 & 1.000 & 0.723 & 0.609 \\
       & 4 & 0.938 & 0.782 & 0.961 & 0.527 & 0.310 \\
\midrule
Zr     & 0 & 0.378 & 0.580 &  0.772 & 0.246 & 0.946 \\
       & 1 & 0.952 & 0.991 & 1.000 & 0.983 & 1.000 \\
       & 2 & 0.532 & 0.707 & 0.994 & 0.374 & 1.000 \\
       & 3 & 0.390 & 0.967 & 1.000 & 0.430 & 1.000 \\
\bottomrule
\end{tabular}
\end{table}

For a more precise statistical view, table \ref{tab:clusterRecallFull} reports, for every HDBSCAN label, how often each conventional detector rediscovers the atoms that the ML workflow has already marked as defective. Here, the ML method is considered as the reference, thus a value of 1.000 means the conventional method never misses a ML outlier in that group; values well below one expose structural motifs the method tends to overlook. Because every \textsc{HDBSCAN} label can now be tied to a specific defect environment as mentioned above (i.e., vacancy cage, dumbbell core, halo atom, large interstitial or vacancy aggregate, see section \ref{ML_res_detect}) the recall values reveal which physical motifs each conventional detector captures or overlooks. 
DXA achieves near‑perfect recall in every well‑defined motif: 1.000 for vacancy cages (FeNiCr‑0, Ni‑2, Zr‑3), dumbbell cores (Ni‑0, FeNiCr‑4) and the isolated interstitial cluster Zr‑1; 0.951 and 0.923 for the large vacancy and interstitial aggregates FeNiCr‑3 and FeNiCr‑1; 0.961 for Ni‑4 and 0.994 for Zr-2.  Its only significant drop is to 0.772 in Zr‑0, a group of strongly distorted interstitial complexes that contains half of all Zr outliers.  That single value explains DXA’s modest reduction in global recall (0.718, cf.\ Table~\ref{tab:PRF}) for the hcp alloy, while its uniform excellence elsewhere underpins the method’s top $F_{1}$ in every system.
CS is virtually indistinguishable from DXA in the two fcc metals: recall $=$ 1.000 in vacancy cages, dumbbell cores and halos (Ni‑0/1, FeNiCr‑4/2) and $\geq0.931$ in the large aggregates (FeNiCr‑3, Ni‑4).  In hcp Zr, however, recall collapses to 0.378–0.532 in groups 0, 2 and 3 that are regions dominated by extended interstitial and vacancy complexes, while remaining 0.952 for the isolated interstitials (Zr‑1).  The sharp symmetry dependence shows that the local centrosymmetry metric cannot distinguish distorted hcp cages from regular ones, erasing a majority of true positives in the alloy with the most interstitial disorder.
PTM behaves as a high‑specificity, template‑limited detector.  It achieves perfect recall (1.000) for dumbbell cores and 0.776–0.804 for vacancy cages in the fcc alloys, but only 0.374–0.430 in Zr aggregates; it retains only 0.115–0.246 of the halo atoms (FeNiCr-2, Ni-1, Zr-0) and 0.444–0.527 of the large interstitial aggregates (FeNiCr-3, Ni-4).  These missing atoms compose roughly half the ML catalogue in each alloy and directly cause PTM’s moderate global recall (0.713–0.494) despite its good precision; any angular distortion beyond the RMSD tolerance invalidates the template match.
CA retains 0.933–0.995 recall in vacancy cages and dumbbell cores but splits in aggregate motifs: 0.683 in the interstitial‑rich FeNiCr‑3 versus 0.782 in the mixed Ni‑4, and a high 0.967 in the vacancy‑rich Zr‑3 yet only 0.580 in the interstitial group Zr‑0.  The pattern indicates that CA over‑counts atoms when the local coordination exceeds 12 (inflating recall in vacancy environments) and begins to miss atoms when compression in an interstitial complex reduces the neighbour count.
Voronoi analysis recalls almost every ML outlier in hcp groups ($\geq$~0.946), including the distorted interstitial complexes of Zr‑0, explaining its high global recall (0.887) but also its poor precision: the same clusters generate many Voronoi‑only alarms.  
In fcc metals recall falls to 0.196–0.716 for vacancy cages and interstitial aggregates.

\subsubsection{Focus on correlation between PTM and ML}

\begin{figure}
	\centering
    \includegraphics[width=0.5\textwidth]{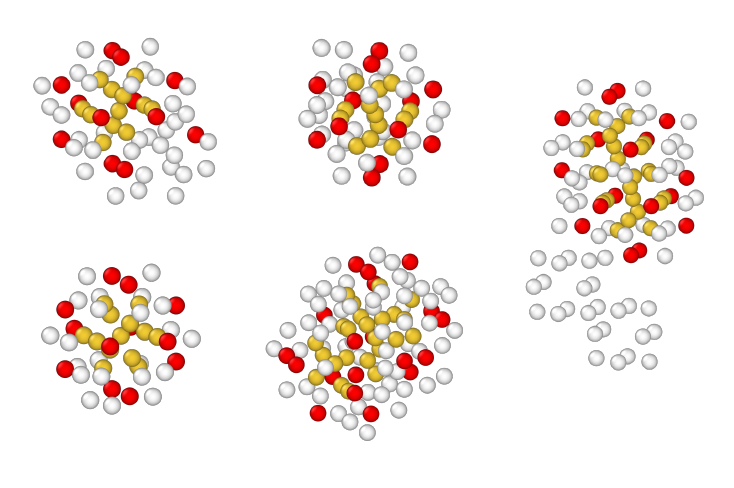}
	\caption{Representative snapshots of the local atomic environments in Ni around atoms identified as ICO (yellow) by PTM surrounded by atoms assigned to HDBSCAN label 3 in red.}
	\label{Fig5}
\end{figure}

The icosahedral template (ICO) in PTM is of particular interest because icosahedra are the basic building blocks of Frank–Kasper phases such as A15 that have been reported as irradiation‑induced defect embryos in fcc metals. Focusing on the PTM‑ICO subset thus offers a test of whether the unsupervised ML pipeline can recognise more complex architecture of such aggregates.
Figure \ref{Fig5} illustrates some examples of this architecture found for Ni. Yellow spheres mark atoms classified as ICO by PTM where icosahedral cages assembled from nearly parallel ⟨100⟩ dumbbells, i.e. A15 kernel, are observed . Red spheres represent atoms assigned to HDBSCAN label~3 by the ML workflow. The images show that group~3 envelops each ICO core with a coherent, albeit distorted, shell.
Quantitatively, only $\sim3\%$ of all ML‑flagged outliers in Ni carry the ICO label, and none of those ICO atoms fall inside HDBSCAN group~3. However, neighbourhood statistics reveal a tight spatial coupling since $\sim86\%$ of group‑3 atoms lie within 4~Å of at least one PTM‑ICO atom, and $\sim87\%$ of PTM‑ICO atoms have a group‑3 neighbour. PTM thus isolates the icosahedral kernel, while the ML pipeline captures the surrounding elastically accommodated shell and merges similar shells into a single descriptor without any structural template. In this sense, the two methods are complementary: PTM provides a chemically intuitive label for the icosahedral core, whereas the data‑driven workflow delineates the extent, thickness, and morphology of the strained envelope and quantifies how far the perturbation propagates into the lattice. Taken together, the figure and the statistics demonstrate that the unsupervised ML pipeline is sensitive to the hierarchical nature of these complex aggregates even though only the innermost icosahedra match a predefined template, thereby validating its usefulness as a template‑agnostic probe of irradiation‑induced defect structures such as fcc Ni.

\section{Conclusion}\label{conclusion}

We have demonstrated that a fully unsupervised, descriptor–based workflow based on SOAP encoding of local atomic environments, autoencoder outlier detection, UMAP dimensionality reduction and HDBSCAN clustering can identify, classify and quantify radiation‑induced defects across chemically and crystallographically distinct systems (FeNiCr, Ni and Zr) from large‐scale 80~keV cascade simulations.  
Although we illustrate the approach using SOAP, any fingerprint that can be cast as a numerical vector or matrix is compatible with the AE–UMAP–HDBSCAN stack. 
The autoencoders isolate fewer than 0.13~\% of the atoms as anomalous, in line with the residual damage expected at this PKA energy, while the subsequent UMAP–HDBSCAN step assigns more than 99.7~\% of those outliers to compact groups that correspond to physically consistent motifs such as monovacancy cages, dumbbell cores, first‑shell "halo" atoms near dumbells, extended vacancy or interstitial aggregates and A15‑like complexes.  A signed cluster‑identification metric based on classical lattice‑site analysis confirms the clean vacancy versus interstitial separation achieved by the pipeline, and second‑order correlations between the number of ML‑flagged atoms per aggregate and its net defect content (\(R^{2}>0.89\)) provide material‑specific calibration curves that translate between atom counts and point‑defect inventory.  Statistical comparison with five conventional detectors shows that our data‑driven catalogue recovers \(70{-}90\%\) of the outliers selected by high‑specificity methods such as DXA and CS while maintaining similarly low false‑positive rates, thereby matching their overall F$_1$ scores yet without template or threshold tuning.  At the same time, the workflow complements template‑limited approaches like PTM by capturing the structurally distorted shells that surround Frank–Kasper A15 structure and other complex clusters. Coupling this ML pipeline with conventional defect metrics yields a powerful hybrid toolkit for comprehensive and interpretable anomaly detection.
Taken together, these results establish the SOAP–AE–UMAP–HDBSCAN efficient framework for quantitative mapping of irradiation defects, offering immediate applications to larger cascade databases and different recoil energies, and enabling anomaly extraction and classification across all types of atomistic simulations beyond irradiation cascades.

\section{Perspectives}\label{perspectives}

The present work opens several directions for future research. First, the workflow will be applied to cascade accumulation in order to follow the progressive build-up of damage under multiple overlapping events. Preliminary tests already indicate that the ML approach not only scales to such datasets but also provides additional insight beyond lattice-site analysis, especially in detecting the overlap and evolution of defect families. Indeed, in these regions, the lattice-site analysis usually fail to provide a reliable identification of defect clusters.
Second, while the unsupervised character of the method is a strength in uncovering unanticipated structures, its reliance on threshold choices, in particular for outlier detection, remains a limitation. Incorporating semi-supervised or hybrid strategies such as integrating a small amount of labelled data or physics-informed constraints  could further stabilize outlier detection and improve interpretability. Such approaches could also enable targeted identification of specific defect patterns by training the model on representative structural motifs (e.g., grain boundaries).
Third, extending the method to other materials and/or different PKA energies will help assess its transferability and reveal structure–property correlations. 

\section{Acknowledgments}

This work was conducted within the framework of the RAISIN project, funded by the NEEDS program. The authors gratefully acknowledge the support of the Centre de Ressources Informatiques (CRI) of the University of Lille.

\appendix{}

\section{Data availability}

The main code and models are available on \href{https://github.com/SamDFr/AE_UMAP_HDBSCAN}{AE\_UMAP\_HDBSCAN} Github repository. MD dataset can be downloaded via Mendeley Data at \cite{NiPd_dataset}. Questions regarding the code or dataset should be addressed to the corresponding author.

\section{List of abbreviations}

\begin{table}[H]
\centering
\caption{List of acronyms used throughout the manuscript.}
\label{tab:acronyms}
\begin{tabular}{ll}
\toprule
Acronym & Definition \\
\midrule
AE & Autoencoder \\
A15 & Frank–Kasper A15 phase (icosahedral defect structure) \\
CA & Coordination Analysis \\
CID & Cluster Identification diagnostic (signed by defect type) \\
CS & Centrosymmetry Parameter \\
CNA & Common Neighbor Analysis \\
DefID & Defect Identification number (cluster size ranking) \\
DXA & Dislocation Extraction Algorithm \\
fcc & Face-Centered Cubic \\
hcp & Hexagonal Close-Packed \\
HDBSCAN & Hierarchical Density-Based Spatial Clustering of Applications with Noise \\
ICO & Icosahedral local environment (PTM template) \\
LAE & Local Atomic Environment \\
MD & Molecular Dynamics \\
ML & Machine Learning \\
MSE & Mean Squared Error (AE reconstruction loss) \\
PKA & Primary Knock-on Atom \\
PTM & Polyhedral Template Matching \\
RMSD & Root Mean Square Deviation \\
SOAP & Smooth Overlap of Atomic Positions (descriptor) \\
TN & True Negative \\
TP & True Positive \\
FN & False Negative \\
FP & False Positive \\
UMAP & Uniform Manifold Approximation and Projection \\
VI & Voronoi Index \\
VOR & Voronoi Analysis \\
\bottomrule
\end{tabular}
\end{table}




\bibliographystyle{cas-model2-names}
\bibliography{cas-refs}

\begin{thebibliography}{71}
\providecommand{\natexlab}[1]{#1}
\providecommand{\url}[1]{\texttt{#1}}
\expandafter\ifx\csname urlstyle\endcsname\relax
  \providecommand{\doi}[1]{doi: #1}\else
  \providecommand{\doi}{doi: \begingroup \urlstyle{rm}\Url}\fi

\bibitem[Liu et~al.(2024)Liu, Lei, and Huang]{liuReviewSynergisticDamage2024}
Hui Liu, Guan-Hong Lei, and He-Fei Huang.
\newblock Review on synergistic damage effect of irradiation and corrosion on
  reactor structural alloys.
\newblock \emph{Nuclear Science and Techniques}, 35\penalty0 (3), March 2024.
\newblock ISSN 2210-3147.
\newblock \doi{10.1007/s41365-024-01415-3}.
\newblock URL \url{http://dx.doi.org/10.1007/s41365-024-01415-3}.

\bibitem[Pomaro(2016)]{pomaroReviewRadiationDamage2016}
Beatrice Pomaro.
\newblock A review on radiation damage in concrete for nuclear facilities: From
  experiments to modeling.
\newblock \emph{Modelling and Simulation in Engineering}, 2016:\penalty0
  1–10, 2016.
\newblock ISSN 1687-5605.
\newblock \doi{10.1155/2016/4165746}.
\newblock URL \url{http://dx.doi.org/10.1155/2016/4165746}.

\bibitem[Zinkle and Busby(2009)]{zinkleStructuralMaterialsFission2009}
Steven~J. Zinkle and Jeremy~T. Busby.
\newblock Structural materials for fission \& fusion energy.
\newblock \emph{Materials Today}, 12\penalty0 (11):\penalty0 12–19, November
  2009.
\newblock ISSN 1369-7021.
\newblock \doi{10.1016/s1369-7021(09)70294-9}.
\newblock URL \url{http://dx.doi.org/10.1016/s1369-7021(09)70294-9}.

\bibitem[Brinkman(1956)]{brinkmanProductionAtomicDisplacements1956}
J.~A. Brinkman.
\newblock Production of atomic displacements by high-energy particles.
\newblock \emph{American Journal of Physics}, 24\penalty0 (4):\penalty0
  246–267, April 1956.
\newblock ISSN 1943-2909.
\newblock \doi{10.1119/1.1934201}.
\newblock URL \url{http://dx.doi.org/10.1119/1.1934201}.

\bibitem[Nordlund et~al.(2018)Nordlund, Zinkle, Sand, Granberg, Averback,
  Stoller, Suzudo, Malerba, Banhart, Weber, Willaime, Dudarev, and
  Simeone]{NORDLUND2018450}
Kai Nordlund, Steven~J. Zinkle, Andrea~E. Sand, Fredric Granberg, Robert~S.
  Averback, Roger~E. Stoller, Tomoaki Suzudo, Lorenzo Malerba, Florian Banhart,
  William~J. Weber, Francois Willaime, Sergei~L. Dudarev, and David Simeone.
\newblock Primary radiation damage: A review of current understanding and
  models.
\newblock \emph{Journal of Nuclear Materials}, 512:\penalty0 450--479, 2018.
\newblock ISSN 0022-3115.
\newblock \doi{https://doi.org/10.1016/j.jnucmat.2018.10.027}.
\newblock URL
  \url{https://www.sciencedirect.com/science/article/pii/S002231151831016X}.

\bibitem[Xiao(2019)]{xiaoFundamentalMechanismsIrradiationHardening2019}
Xiazi Xiao.
\newblock Fundamental mechanisms for irradiation-hardening and embrittlement: A
  review.
\newblock \emph{Metals}, 9\penalty0 (10):\penalty0 1132, October 2019.
\newblock ISSN 2075-4701.
\newblock \doi{10.3390/met9101132}.
\newblock URL \url{http://dx.doi.org/10.3390/met9101132}.

\bibitem[Reali et~al.(2025)Reali, Boleininger, Mason, and
  Dudarev]{realiAtomisticSimulationsAthermal2025}
Luca Reali, Max Boleininger, Daniel~R. Mason, and Sergei~L. Dudarev.
\newblock Atomistic simulations of athermal irradiation creep and swelling of
  copper and tungsten at high dose.
\newblock \emph{Acta Materialia}, 288:\penalty0 120814, April 2025.
\newblock ISSN 1359-6454.
\newblock \doi{10.1016/j.actamat.2025.120814}.
\newblock URL \url{http://dx.doi.org/10.1016/j.actamat.2025.120814}.

\bibitem[Nordlund(2019)]{nordlundHistoricalReviewComputer2019}
K.~Nordlund.
\newblock Historical review of computer simulation of radiation effects in
  materials.
\newblock \emph{Journal of Nuclear Materials}, 520:\penalty0 273–295, July
  2019.
\newblock ISSN 0022-3115.
\newblock \doi{10.1016/j.jnucmat.2019.04.028}.
\newblock URL \url{http://dx.doi.org/10.1016/j.jnucmat.2019.04.028}.

\bibitem[Stoller(2012)]{stoller111PrimaryRadiation}
R.E. Stoller.
\newblock \emph{Primary Radiation Damage Formation}, page 293–332.
\newblock Elsevier, 2012.
\newblock ISBN 9780080560335.
\newblock \doi{10.1016/b978-0-08-056033-5.00027-6}.
\newblock URL \url{http://dx.doi.org/10.1016/b978-0-08-056033-5.00027-6}.

\bibitem[Becquart et~al.()Becquart, De~Backer, and
  Domain]{becquartAtomisticModelingRadiation2018}
Charlotte~S. Becquart, Andrée De~Backer, and Christophe Domain.
\newblock Atomistic {{Modeling}} of {{Radiation Damage}} in {{Metallic
  Alloys}}.
\newblock In Siegfried Schmauder, Chuin-Shan Chen, Krishan~K. Chawla, Nikhilesh
  Chawla, Weiqiu Chen, and Yutaka Kagawa, editors, \emph{Handbook of
  {{Mechanics}} of {{Materials}}}, pages 1--30. Springer Singapore.
\newblock ISBN 978-981-10-6855-3.
\newblock \doi{10.1007/978-981-10-6855-3_21-1}.

\bibitem[Becquart and
  Domain(2010)]{becquartModelingMicrostructureIrradiation2011}
C.~S. Becquart and C.~Domain.
\newblock Modeling microstructure and irradiation effects.
\newblock \emph{Metallurgical and Materials Transactions A}, 42\penalty0
  (4):\penalty0 852–870, December 2010.
\newblock ISSN 1543-1940.
\newblock \doi{10.1007/s11661-010-0460-7}.
\newblock URL \url{http://dx.doi.org/10.1007/s11661-010-0460-7}.

\bibitem[Kelchner et~al.(1998)Kelchner, Plimpton, and
  Hamilton]{kelchnerDislocationNucleationDefect1998}
Cynthia~L. Kelchner, S.~J. Plimpton, and J.~C. Hamilton.
\newblock Dislocation nucleation and defect structure during surface
  indentation.
\newblock \emph{Physical Review B}, 58\penalty0 (17):\penalty0 11085–11088,
  November 1998.
\newblock ISSN 1095-3795.
\newblock \doi{10.1103/physrevb.58.11085}.
\newblock URL \url{http://dx.doi.org/10.1103/physrevb.58.11085}.

\bibitem[Honeycutt and Andersen(1987)]{honeycuttMolecularDynamicsStudy1987}
J.~Dana. Honeycutt and Hans~C. Andersen.
\newblock Molecular dynamics study of melting and freezing of small
  lennard-jones clusters.
\newblock \emph{The Journal of Physical Chemistry}, 91\penalty0 (19):\penalty0
  4950–4963, September 1987.
\newblock ISSN 1541-5740.
\newblock \doi{10.1021/j100303a014}.
\newblock URL \url{http://dx.doi.org/10.1021/j100303a014}.

\bibitem[Faken and Jónsson(1994)]{fakenSystematicAnalysisLocal1994}
Daniel Faken and Hannes Jónsson.
\newblock Systematic analysis of local atomic structure combined with 3d
  computer graphics.
\newblock \emph{Computational Materials Science}, 2\penalty0 (2):\penalty0
  279–286, March 1994.
\newblock ISSN 0927-0256.
\newblock \doi{10.1016/0927-0256(94)90109-0}.
\newblock URL \url{http://dx.doi.org/10.1016/0927-0256(94)90109-0}.

\bibitem[Malins et~al.(2013)Malins, Williams, Eggers, and
  Royall]{malinsIdentificationStructureCondensed2013}
Alex Malins, Stephen~R. Williams, Jens Eggers, and C.~Patrick Royall.
\newblock Identification of structure in condensed matter with the topological
  cluster classification.
\newblock \emph{The Journal of Chemical Physics}, 139\penalty0 (23), December
  2013.
\newblock ISSN 1089-7690.
\newblock \doi{10.1063/1.4832897}.
\newblock URL \url{http://dx.doi.org/10.1063/1.4832897}.

\bibitem[Larsen et~al.(2016)Larsen, Schmidt, and
  Schiøtz]{larsenRobustStructuralIdentification2016}
Peter~Mahler Larsen, Søren Schmidt, and Jakob Schiøtz.
\newblock Robust structural identification via polyhedral template matching.
\newblock \emph{Modelling and Simulation in Materials Science and Engineering},
  24\penalty0 (5):\penalty0 055007, May 2016.
\newblock ISSN 1361-651X.
\newblock \doi{10.1088/0965-0393/24/5/055007}.
\newblock URL \url{http://dx.doi.org/10.1088/0965-0393/24/5/055007}.

\bibitem[Stukowski and
  Albe(2010)]{stukowskiExtractingDislocationsNondislocation2010}
Alexander Stukowski and Karsten Albe.
\newblock Extracting dislocations and non-dislocation crystal defects from
  atomistic simulation data.
\newblock \emph{Modelling and Simulation in Materials Science and Engineering},
  18\penalty0 (8):\penalty0 085001, September 2010.
\newblock ISSN 1361-651X.
\newblock \doi{10.1088/0965-0393/18/8/085001}.
\newblock URL \url{http://dx.doi.org/10.1088/0965-0393/18/8/085001}.

\bibitem[Stukowski et~al.(2012)Stukowski, Bulatov, and
  Arsenlis]{stukowskiAutomatedIdentificationIndexing2012}
Alexander Stukowski, Vasily~V Bulatov, and Athanasios Arsenlis.
\newblock Automated identification and indexing of dislocations in crystal
  interfaces.
\newblock \emph{Modelling and Simulation in Materials Science and Engineering},
  20\penalty0 (8):\penalty0 085007, October 2012.
\newblock ISSN 1361-651X.
\newblock \doi{10.1088/0965-0393/20/8/085007}.
\newblock URL \url{http://dx.doi.org/10.1088/0965-0393/20/8/085007}.

\bibitem[Lazar et~al.(2015)Lazar, Han, and
  Srolovitz]{lazarTopologicalFrameworkLocal2015a}
Emanuel~A. Lazar, Jian Han, and David~J. Srolovitz.
\newblock Topological framework for local structure analysis in condensed
  matter.
\newblock \emph{Proceedings of the National Academy of Sciences}, 112\penalty0
  (43), October 2015.
\newblock ISSN 1091-6490.
\newblock \doi{10.1073/pnas.1505788112}.
\newblock URL \url{http://dx.doi.org/10.1073/pnas.1505788112}.

\bibitem[De~Backer et~al.(2021)De~Backer, Becquart, Olsson, and
  Domain]{debackerModellingPrimaryDamage2021}
Andrée De~Backer, Charlotte~S. Becquart, Pär Olsson, and Christophe Domain.
\newblock Modelling the primary damage in fe and w: influence of the
  short-range interactions on the cascade properties: Part 2 – multivariate
  multiple linear regression analysis of displacement cascades.
\newblock \emph{Journal of Nuclear Materials}, 549:\penalty0 152887, June 2021.
\newblock ISSN 0022-3115.
\newblock \doi{10.1016/j.jnucmat.2021.152887}.
\newblock URL \url{http://dx.doi.org/10.1016/j.jnucmat.2021.152887}.

\bibitem[Möller and Bitzek(2016)]{mollerBDANovelMethod2016}
Johannes~J. Möller and Erik Bitzek.
\newblock Bda: A novel method for identifying defects in body-centered cubic
  crystals.
\newblock \emph{MethodsX}, 3:\penalty0 279–288, 2016.
\newblock ISSN 2215-0161.
\newblock \doi{10.1016/j.mex.2016.03.013}.
\newblock URL \url{http://dx.doi.org/10.1016/j.mex.2016.03.013}.

\bibitem[Snow et~al.(2019)Snow, Doty, and
  Johnson]{snowSimpleApproachAtomic2019}
Brandon~D. Snow, Dustin~D. Doty, and Oliver~K. Johnson.
\newblock A simple approach to atomic structure characterization for machine
  learning of grain boundary structure-property models.
\newblock \emph{Frontiers in Materials}, 6, May 2019.
\newblock ISSN 2296-8016.
\newblock \doi{10.3389/fmats.2019.00120}.
\newblock URL \url{http://dx.doi.org/10.3389/fmats.2019.00120}.

\bibitem[Furstoss et~al.(2025)Furstoss, Salazar, Carrez, Hirel, and
  Lam]{furstossAllaroundLocalStructure2025}
Jean Furstoss, Carlos~R. Salazar, Philippe Carrez, Pierre Hirel, and Julien
  Lam.
\newblock All-around local structure classification with supervised learning:
  The example of crystal phases and dislocations in complex oxides.
\newblock \emph{Computer Physics Communications}, 309:\penalty0 109480, April
  2025.
\newblock ISSN 0010-4655.
\newblock \doi{10.1016/j.cpc.2024.109480}.
\newblock URL \url{http://dx.doi.org/10.1016/j.cpc.2024.109480}.

\bibitem[Bartók et~al.(2013)Bartók, Kondor, and
  Csányi]{bartokRepresentingChemicalEnvironments2013}
Albert~P. Bartók, Risi Kondor, and Gábor Csányi.
\newblock On representing chemical environments.
\newblock \emph{Physical Review B}, 87\penalty0 (18), May 2013.
\newblock ISSN 1550-235X.
\newblock \doi{10.1103/physrevb.87.184115}.
\newblock URL \url{http://dx.doi.org/10.1103/physrevb.87.184115}.

\bibitem[Steinhardt et~al.(1983)Steinhardt, Nelson, and
  Ronchetti]{steinhardtBondorientationalOrderLiquids1983}
Paul~J. Steinhardt, David~R. Nelson, and Marco Ronchetti.
\newblock Bond-orientational order in liquids and glasses.
\newblock \emph{Physical Review B}, 28\penalty0 (2):\penalty0 784–805, July
  1983.
\newblock ISSN 0163-1829.
\newblock \doi{10.1103/physrevb.28.784}.
\newblock URL \url{http://dx.doi.org/10.1103/physrevb.28.784}.

\bibitem[Lechner and Dellago(2008)]{lechnerAccurateDeterminationCrystal2008}
Wolfgang Lechner and Christoph Dellago.
\newblock Accurate determination of crystal structures based on averaged local
  bond order parameters.
\newblock \emph{The Journal of Chemical Physics}, 129\penalty0 (11), September
  2008.
\newblock ISSN 1089-7690.
\newblock \doi{10.1063/1.2977970}.
\newblock URL \url{http://dx.doi.org/10.1063/1.2977970}.

\bibitem[Behler and
  Parrinello(2007)]{behlerGeneralizedNeuralNetworkRepresentation2007}
Jörg Behler and Michele Parrinello.
\newblock Generalized neural-network representation of high-dimensional
  potential-energy surfaces.
\newblock \emph{Physical Review Letters}, 98\penalty0 (14), April 2007.
\newblock ISSN 1079-7114.
\newblock \doi{10.1103/physrevlett.98.146401}.
\newblock URL \url{http://dx.doi.org/10.1103/physrevlett.98.146401}.

\bibitem[Behler(2011)]{behlerAtomcenteredSymmetryFunctions2011a}
Jörg Behler.
\newblock Atom-centered symmetry functions for constructing high-dimensional
  neural network potentials.
\newblock \emph{The Journal of Chemical Physics}, 134\penalty0 (7), February
  2011.
\newblock ISSN 1089-7690.
\newblock \doi{10.1063/1.3553717}.
\newblock URL \url{http://dx.doi.org/10.1063/1.3553717}.

\bibitem[Allera et~al.(2024)Allera, Goryaeva, Lafourcade, Maillet, and
  Marinica]{alleraNeighborsMapEfficient2024}
Arnaud Allera, Alexandra~M. Goryaeva, Paul Lafourcade, Jean-Bernard Maillet,
  and Mihai-Cosmin Marinica.
\newblock Neighbors map: An efficient atomic descriptor for structural
  analysis.
\newblock \emph{Computational Materials Science}, 231:\penalty0 112535, January
  2024.
\newblock ISSN 0927-0256.
\newblock \doi{10.1016/j.commatsci.2023.112535}.
\newblock URL \url{http://dx.doi.org/10.1016/j.commatsci.2023.112535}.

\bibitem[Banik et~al.(2023)Banik, Dhabal, Chan, Manna, Cherukara, Molinero, and
  Sankaranarayanan]{banikCEGANNCrystalEdge2023}
Suvo Banik, Debdas Dhabal, Henry Chan, Sukriti Manna, Mathew Cherukara, Valeria
  Molinero, and Subramanian K. R.~S. Sankaranarayanan.
\newblock Cegann: Crystal edge graph attention neural network for multiscale
  classification of materials environment.
\newblock \emph{npj Computational Materials}, 9\penalty0 (1), February 2023.
\newblock ISSN 2057-3960.
\newblock \doi{10.1038/s41524-023-00975-z}.
\newblock URL \url{http://dx.doi.org/10.1038/s41524-023-00975-z}.

\bibitem[Goryaeva et~al.(2020)Goryaeva, Lapointe, Dai, Dérès, Maillet, and
  Marinica]{goryaevaReinforcingMaterialsModelling2020}
Alexandra~M. Goryaeva, Clovis Lapointe, Chendi Dai, Julien Dérès,
  Jean-Bernard Maillet, and Mihai-Cosmin Marinica.
\newblock Reinforcing materials modelling by encoding the structures of defects
  in crystalline solids into distortion scores.
\newblock \emph{Nature Communications}, 11\penalty0 (1), September 2020.
\newblock ISSN 2041-1723.
\newblock \doi{10.1038/s41467-020-18282-2}.
\newblock URL \url{http://dx.doi.org/10.1038/s41467-020-18282-2}.

\bibitem[von Toussaint et~al.(2021)von Toussaint, Domínguez-Gutiérrez,
  Compostella, and Rampp]{vontoussaintFaVADSoftwareWorkflow2021}
Udo von Toussaint, F.J. Domínguez-Gutiérrez, Michele Compostella, and Markus
  Rampp.
\newblock Favad: A software workflow for characterization and visualizing of
  defects in crystalline structures.
\newblock \emph{Computer Physics Communications}, 262:\penalty0 107816, May
  2021.
\newblock ISSN 0010-4655.
\newblock \doi{10.1016/j.cpc.2020.107816}.
\newblock URL \url{http://dx.doi.org/10.1016/j.cpc.2020.107816}.

\bibitem[Becker et~al.(2022)Becker, Devijver, Molinier, and
  Jakse]{beckerUnsupervisedTopologicalLearning2022}
Sébastien Becker, Emilie Devijver, Rémi Molinier, and Noël Jakse.
\newblock Unsupervised topological learning for identification of atomic
  structures.
\newblock \emph{Physical Review E}, 105\penalty0 (4), April 2022.
\newblock ISSN 2470-0053.
\newblock \doi{10.1103/physreve.105.045304}.
\newblock URL \url{http://dx.doi.org/10.1103/physreve.105.045304}.

\bibitem[Geiger and Dellago(2013)]{geigerNeuralNetworksLocal2013}
Philipp Geiger and Christoph Dellago.
\newblock Neural networks for local structure detection in polymorphic systems.
\newblock \emph{The Journal of Chemical Physics}, 139\penalty0 (16), October
  2013.
\newblock ISSN 1089-7690.
\newblock \doi{10.1063/1.4825111}.
\newblock URL \url{http://dx.doi.org/10.1063/1.4825111}.

\bibitem[Chung et~al.(2022)Chung, Freitas, Cheon, and
  Reed]{chungDatacentricFrameworkCrystal2022}
Heejung~W. Chung, Rodrigo Freitas, Gowoon Cheon, and Evan~J. Reed.
\newblock Data-centric framework for crystal structure identification in
  atomistic simulations using machine learning.
\newblock \emph{Physical Review Materials}, 6\penalty0 (4), April 2022.
\newblock ISSN 2475-9953.
\newblock \doi{10.1103/physrevmaterials.6.043801}.
\newblock URL \url{http://dx.doi.org/10.1103/physrevmaterials.6.043801}.

\bibitem[Ziletti et~al.(2018)Ziletti, Kumar, Scheffler, and
  Ghiringhelli]{zilettiInsightfulClassificationCrystal2018}
Angelo Ziletti, Devinder Kumar, Matthias Scheffler, and Luca~M. Ghiringhelli.
\newblock Insightful classification of crystal structures using deep learning.
\newblock \emph{Nature Communications}, 9\penalty0 (1), July 2018.
\newblock ISSN 2041-1723.
\newblock \doi{10.1038/s41467-018-05169-6}.
\newblock URL \url{http://dx.doi.org/10.1038/s41467-018-05169-6}.

\bibitem[Leitherer et~al.(2021)Leitherer, Ziletti, and
  Ghiringhelli]{leithererRobustRecognitionExploratory2021}
Andreas Leitherer, Angelo Ziletti, and Luca~M. Ghiringhelli.
\newblock Robust recognition and exploratory analysis of crystal structures via
  bayesian deep learning.
\newblock \emph{Nature Communications}, 12\penalty0 (1), October 2021.
\newblock ISSN 2041-1723.
\newblock \doi{10.1038/s41467-021-26511-5}.
\newblock URL \url{http://dx.doi.org/10.1038/s41467-021-26511-5}.

\bibitem[Boattini et~al.(2019)Boattini, Dijkstra, and
  Filion]{boattiniUnsupervisedLearningLocal2019}
Emanuele Boattini, Marjolein Dijkstra, and Laura Filion.
\newblock Unsupervised learning for local structure detection in colloidal
  systems.
\newblock \emph{The Journal of Chemical Physics}, 151\penalty0 (15), October
  2019.
\newblock ISSN 1089-7690.
\newblock \doi{10.1063/1.5118867}.
\newblock URL \url{http://dx.doi.org/10.1063/1.5118867}.

\bibitem[Zhang et~al.(2023)Zhang, Yao, Li, Python, and
  Mochizuki]{zhangFastCrystalGrowth2023}
Xuan Zhang, Yifeng Yao, Hongyi Li, Andre Python, and Kenji Mochizuki.
\newblock Fast crystal growth of ice vii owing to the decoupling of
  translational and rotational ordering.
\newblock \emph{Communications Physics}, 6\penalty0 (1), July 2023.
\newblock ISSN 2399-3650.
\newblock \doi{10.1038/s42005-023-01285-y}.
\newblock URL \url{http://dx.doi.org/10.1038/s42005-023-01285-y}.

\bibitem[Berahmand et~al.(2024)Berahmand, Daneshfar, Salehi, Li, and
  Xu]{berahmandAutoencodersTheirApplications2024}
Kamal Berahmand, Fatemeh Daneshfar, Elaheh~Sadat Salehi, Yuefeng Li, and Yue
  Xu.
\newblock Autoencoders and their applications in machine learning: a survey.
\newblock \emph{Artificial Intelligence Review}, 57\penalty0 (2), February
  2024.
\newblock ISSN 1573-7462.
\newblock \doi{10.1007/s10462-023-10662-6}.
\newblock URL \url{http://dx.doi.org/10.1007/s10462-023-10662-6}.

\bibitem[Bhardwaj et~al.(2020)Bhardwaj, Sand, and
  Warrier]{bhardwajClassificationClustersCollision2020}
Utkarsh Bhardwaj, Andrea~E. Sand, and Manoj Warrier.
\newblock Classification of clusters in collision cascades.
\newblock \emph{Computational Materials Science}, 172:\penalty0 109364,
  February 2020.
\newblock ISSN 0927-0256.
\newblock \doi{10.1016/j.commatsci.2019.109364}.
\newblock URL \url{http://dx.doi.org/10.1016/j.commatsci.2019.109364}.

\bibitem[Campello et~al.(2015)Campello, Moulavi, Zimek, and
  Sander]{campelloHierarchicalDensityEstimates2015}
Ricardo J. G.~B. Campello, Davoud Moulavi, Arthur Zimek, and Jörg Sander.
\newblock Hierarchical density estimates for data clustering, visualization,
  and outlier detection.
\newblock \emph{ACM Transactions on Knowledge Discovery from Data}, 10\penalty0
  (1):\penalty0 1–51, July 2015.
\newblock ISSN 1556-472X.
\newblock \doi{10.1145/2733381}.
\newblock URL \url{http://dx.doi.org/10.1145/2733381}.

\bibitem[McInnes et~al.()McInnes, Healy, and
  Melville]{mcinnesUMAPUniformManifold2018}
Leland McInnes, John Healy, and James Melville.
\newblock {{UMAP}}: {{Uniform Manifold Approximation}} and {{Projection}} for
  {{Dimension Reduction}}.

\bibitem[Roncoroni et~al.(2023)Roncoroni, Sanz-Matias, Sundararaman, and
  Prendergast]{roncoroniUnsupervisedLearningRepresentative2023}
Fabrice Roncoroni, Ana Sanz-Matias, Siddharth Sundararaman, and David
  Prendergast.
\newblock Unsupervised learning of representative local atomic arrangements in
  molecular dynamics data.
\newblock \emph{Physical Chemistry Chemical Physics}, 25\penalty0
  (19):\penalty0 13741–13754, 2023.
\newblock ISSN 1463-9084.
\newblock \doi{10.1039/d3cp00525a}.
\newblock URL \url{http://dx.doi.org/10.1039/d3cp00525a}.

\bibitem[Kývala et~al.(2025)Kývala, Montero~de Hijes, and
  Dellago]{kyvalaUnsupervisedIdentificationCrystal2025}
Lukáš Kývala, Pablo Montero~de Hijes, and Christoph Dellago.
\newblock Unsupervised identification of crystal defects from atomistic
  potential descriptors.
\newblock \emph{npj Computational Materials}, 11\penalty0 (1), February 2025.
\newblock ISSN 2057-3960.
\newblock \doi{10.1038/s41524-025-01544-2}.
\newblock URL \url{http://dx.doi.org/10.1038/s41524-025-01544-2}.

\bibitem[Maćkiewicz and
  Ratajczak(1993)]{mackiewiczPrincipalComponentsAnalysis1993}
Andrzej Maćkiewicz and Waldemar Ratajczak.
\newblock Principal components analysis (pca).
\newblock \emph{Computers \& Geosciences}, 19\penalty0 (3):\penalty0 303–342,
  March 1993.
\newblock ISSN 0098-3004.
\newblock \doi{10.1016/0098-3004(93)90090-r}.
\newblock URL \url{http://dx.doi.org/10.1016/0098-3004(93)90090-r}.

\bibitem[van~der Maaten and Hinton(2008)]{JMLR:v9:vandermaaten08a}
Laurens van~der Maaten and Geoffrey Hinton.
\newblock Visualizing data using t-sne.
\newblock \emph{Journal of Machine Learning Research}, 9\penalty0
  (86):\penalty0 2579--2605, 2008.
\newblock URL \url{http://jmlr.org/papers/v9/vandermaaten08a.html}.

\bibitem[Allaoui et~al.(2020)Allaoui, Kherfi, and
  Cheriet]{allaouiConsiderablyImprovingClustering2020}
Mebarka Allaoui, Mohammed~Lamine Kherfi, and Abdelhakim Cheriet.
\newblock \emph{Considerably Improving Clustering Algorithms Using UMAP
  Dimensionality Reduction Technique: A Comparative Study}, page 317–325.
\newblock Springer International Publishing, 2020.
\newblock ISBN 9783030519353.
\newblock \doi{10.1007/978-3-030-51935-3_34}.
\newblock URL \url{http://dx.doi.org/10.1007/978-3-030-51935-3_34}.

\bibitem[Herrmann et~al.(2023)Herrmann, Kazempour, Scheipl, and
  Kröger]{herrmannEnhancingClusterAnalysis2024}
Moritz Herrmann, Daniyal Kazempour, Fabian Scheipl, and Peer Kröger.
\newblock Enhancing cluster analysis via topological manifold learning.
\newblock \emph{Data Mining and Knowledge Discovery}, 38\penalty0 (3):\penalty0
  840–887, September 2023.
\newblock ISSN 1573-756X.
\newblock \doi{10.1007/s10618-023-00980-2}.
\newblock URL \url{http://dx.doi.org/10.1007/s10618-023-00980-2}.

\bibitem[Nair et~al.(2025)Nair, Becquart, Domain, and
  De~Backer]{nairStatisticalStudyDisplacement2025}
Adithya Nair, Charlotte~S. Becquart, Christophe Domain, and Andrée De~Backer.
\newblock Statistical study of displacement cascades in ni and fenicr alloys:
  Understanding the influence of potential and composition on primary damage
  modeling.
\newblock \emph{Journal of Nuclear Materials}, 612:\penalty0 155832, June 2025.
\newblock ISSN 0022-3115.
\newblock \doi{10.1016/j.jnucmat.2025.155832}.
\newblock URL \url{http://dx.doi.org/10.1016/j.jnucmat.2025.155832}.

\bibitem[Becquart et~al.(1997)Becquart, Decker, Domain, Ruste, Souffez,
  Turbatte, and Van~Duysen]{DYMOKA}
C.~S. Becquart, K.~M. Decker, C.~Domain, J.~Ruste, Y.~Souffez, J.~C. Turbatte,
  and J.~C. Van~Duysen.
\newblock Massively parallel molecular dynamics simulations with eam
  potentials.
\newblock \emph{Radiation Effects and Defects in Solids}, 142\penalty0
  (1–4):\penalty0 9–21, June 1997.
\newblock ISSN 1029-4953.
\newblock \doi{10.1080/10420159708211592}.
\newblock URL \url{http://dx.doi.org/10.1080/10420159708211592}.

\bibitem[Bonny et~al.(2011)Bonny, Terentyev, Pasianot, Poncé, and
  Bakaev]{Bonny2011}
G~Bonny, D~Terentyev, R~C Pasianot, S~Poncé, and A~Bakaev.
\newblock Interatomic potential to study plasticity in stainless steels: the
  fenicr model alloy.
\newblock \emph{Modelling and Simulation in Materials Science and Engineering},
  19\penalty0 (8):\penalty0 085008, nov 2011.
\newblock \doi{10.1088/0965-0393/19/8/085008}.
\newblock URL \url{https://dx.doi.org/10.1088/0965-0393/19/8/085008}.

\bibitem[Béland et~al.(2017)Béland, Tamm, Mu, Samolyuk, Osetsky, Aabloo,
  Klintenberg, Caro, and Stoller]{Beland2017}
Laurent~Karim Béland, Artur Tamm, Sai Mu, German D. Samolyuk, Yuri N.
  Osetsky, Alvo Aabloo, Mattias Klintenberg, Alfredo Caro, and Roger E.
  Stoller.
\newblock Accurate classical short-range forces for the study of collision
  cascades in fe–ni–cr.
\newblock \emph{Computer Physics Communications}, 219:\penalty0 11--19, 2017.
\newblock ISSN 0010-4655.
\newblock \doi{https://doi.org/10.1016/j.cpc.2017.05.001}.
\newblock URL
  \url{https://www.sciencedirect.com/science/article/pii/S0010465517301315}.

\bibitem[Mishin(2004)]{Mishin2004}
Y.~Mishin.
\newblock Atomistic modeling of the $\gamma$ and $\gamma'$-phases of the ni-al
  system.
\newblock \emph{Acta Materialia}, 52\penalty0 (6):\penalty0 1451--1467, 2004.
\newblock ISSN 1359-6454.
\newblock \doi{https://doi.org/10.1016/j.actamat.2003.11.026}.
\newblock URL
  \url{https://www.sciencedirect.com/science/article/pii/S1359645403007262}.

\bibitem[Samolyuk et~al.(2016)Samolyuk, Béland, Stocks, and
  Stoller]{Samolyuk2016}
G~D Samolyuk, L~K Béland, G~M Stocks, and R~E Stoller.
\newblock Electron–phonon coupling in ni-based binary alloys with application
  to displacement cascade modeling.
\newblock \emph{Journal of Physics: Condensed Matter}, 28\penalty0
  (17):\penalty0 175501, apr 2016.
\newblock \doi{10.1088/0953-8984/28/17/175501}.
\newblock URL \url{https://dx.doi.org/10.1088/0953-8984/28/17/175501}.

\bibitem[Mendelev and Ackland(2007)]{Mendelev2007}
M.~I. Mendelev and G.~J. Ackland.
\newblock Development of an interatomic potential for the simulation of phase
  transformations in zirconium.
\newblock \emph{Philosophical Magazine Letters}, 87\penalty0 (5):\penalty0
  349--359, 2007.
\newblock \doi{10.1080/09500830701191393}.
\newblock URL \url{https://doi.org/10.1080/09500830701191393}.

\bibitem[Stoller et~al.(1997)Stoller, Odette, and
  Wirth]{stollerPrimaryDamageFormation1997}
R.E. Stoller, G.R. Odette, and B.D. Wirth.
\newblock Primary damage formation in bcc iron.
\newblock \emph{Journal of Nuclear Materials}, 251:\penalty0 49–60, November
  1997.
\newblock ISSN 0022-3115.
\newblock \doi{10.1016/s0022-3115(97)00256-0}.
\newblock URL \url{http://dx.doi.org/10.1016/s0022-3115(97)00256-0}.

\bibitem[Becquart et~al.(2021)Becquart, De~Backer, Olsson, and
  Domain]{becquartModellingPrimaryDamage2021}
Charlotte~S. Becquart, Andr{\'e}e De~Backer, P{\"a}r Olsson, and Christophe
  Domain.
\newblock Modelling the primary damage in {{Fe}} and {{W}}: {{Influence}} of
  the short range interactions on the cascade properties: {{Part}} 1 --
  {{Energy}} transfer.
\newblock \emph{Journal of Nuclear Materials}, 547:\penalty0 152816, April
  2021.
\newblock ISSN 00223115.
\newblock \doi{10.1016/j.jnucmat.2021.152816}.

\bibitem[A.~De~Backer(2025)]{Debacker_zr2025}
A.~Legris L.~Thuinet A.~De~Backer, C.~Domain.
\newblock Displacement cascades in zr and zrnb alloys.
\newblock \emph{To be published}, 2025.

\bibitem[Himanen et~al.(2020)Himanen, J{\"a}ger, Morooka, Federici~Canova,
  Ranawat, Gao, Rinke, and Foster]{dscribe}
Lauri Himanen, Marc O.~J. J{\"a}ger, Eiaki~V. Morooka, Filippo Federici~Canova,
  Yashasvi~S. Ranawat, David~Z. Gao, Patrick Rinke, and Adam~S. Foster.
\newblock {DScribe: Library of descriptors for machine learning in materials
  science}.
\newblock \emph{Computer Physics Communications}, 247:\penalty0 106949, 2020.
\newblock ISSN 0010-4655.
\newblock \doi{10.1016/j.cpc.2019.106949}.
\newblock URL \url{https://doi.org/10.1016/j.cpc.2019.106949}.

\bibitem[Laakso et~al.(2023)Laakso, Himanen, Homm, Morooka, J{\"a}ger,
  Todorovi{\'c}, and Rinke]{dscribe2}
Jarno Laakso, Lauri Himanen, Henrietta Homm, Eiaki~V Morooka, Marc~OJ
  J{\"a}ger, Milica Todorovi{\'c}, and Patrick Rinke.
\newblock Updates to the dscribe library: New descriptors and derivatives.
\newblock \emph{The Journal of Chemical Physics}, 158\penalty0 (23), 2023.

\bibitem[Pedregosa et~al.(2011)Pedregosa, Varoquaux, Gramfort, Michel, Thirion,
  Grisel, Blondel, Prettenhofer, Weiss, Dubourg, Vanderplas, Passos,
  Cournapeau, Brucher, Perrot, and Duchesnay]{scikit-learn}
F.~Pedregosa, G.~Varoquaux, A.~Gramfort, V.~Michel, B.~Thirion, O.~Grisel,
  M.~Blondel, P.~Prettenhofer, R.~Weiss, V.~Dubourg, J.~Vanderplas, A.~Passos,
  D.~Cournapeau, M.~Brucher, M.~Perrot, and E.~Duchesnay.
\newblock Scikit-learn: Machine learning in {P}ython.
\newblock \emph{Journal of Machine Learning Research}, 12:\penalty0 2825--2830,
  2011.

\bibitem[Paszke et~al.(2019)Paszke, Gross, Massa, Lerer, Bradbury, Chanan,
  Killeen, Lin, Gimelshein, Antiga, Desmaison, Kopf, Yang, DeVito, Raison,
  Tejani, Chilamkurthy, Steiner, Fang, Bai, and Chintala]{NEURIPS2019_bdbca288}
Adam Paszke, Sam Gross, Francisco Massa, Adam Lerer, James Bradbury, Gregory
  Chanan, Trevor Killeen, Zeming Lin, Natalia Gimelshein, Luca Antiga, Alban
  Desmaison, Andreas Kopf, Edward Yang, Zachary DeVito, Martin Raison, Alykhan
  Tejani, Sasank Chilamkurthy, Benoit Steiner, Lu~Fang, Junjie Bai, and Soumith
  Chintala.
\newblock Pytorch: An imperative style, high-performance deep learning library.
\newblock In H.~Wallach, H.~Larochelle, A.~Beygelzimer, F.~d\textquotesingle
  Alch\'{e}-Buc, E.~Fox, and R.~Garnett, editors, \emph{Advances in Neural
  Information Processing Systems}, volume~32. Curran Associates, Inc., 2019.
\newblock URL
  \url{https://proceedings.neurips.cc/paper_files/paper/2019/file/bdbca288fee7f92f2bfa9f7012727740-Paper.pdf}.

\bibitem[Akiba et~al.()Akiba, Sano, Yanase, Ohta, and
  Koyama]{akibaOptunaNextgenerationHyperparameter2019}
Takuya Akiba, Shotaro Sano, Toshihiko Yanase, Takeru Ohta, and Masanori Koyama.
\newblock Optuna: {{A Next-generation Hyperparameter Optimization Framework}}.
\newblock In \emph{Proceedings of the 25th {{ACM SIGKDD International
  Conference}} on {{Knowledge Discovery}} \& {{Data Mining}}}, pages
  2623--2631. ACM.
\newblock ISBN 978-1-4503-6201-6.
\newblock \doi{10.1145/3292500.3330701}.

\bibitem[Kingma and Ba()]{kingmaAdamMethodStochastic2014}
Diederik~P. Kingma and Jimmy Ba.
\newblock Adam: {{A Method}} for {{Stochastic Optimization}}.
\newblock URL \url{https://doi.org/10.48550/arXiv.1412.6980}.

\bibitem[De~Backer et~al.(2018)De~Backer, Domain, Becquart, Luneville, Simeone,
  Sand, and Nordlund]{debackerModelDefectCluster2018}
A~De~Backer, C~Domain, C~S Becquart, L~Luneville, D~Simeone, A~E Sand, and
  K~Nordlund.
\newblock A model of defect cluster creation in fragmented cascades in metals
  based on morphological analysis.
\newblock \emph{Journal of Physics: Condensed Matter}, 30\penalty0
  (40):\penalty0 405701, September 2018.
\newblock ISSN 1361-648X.
\newblock \doi{10.1088/1361-648x/aadb4e}.
\newblock URL \url{http://dx.doi.org/10.1088/1361-648x/aadb4e}.

\bibitem[Stukowski(2009)]{ovito}
Alexander Stukowski.
\newblock Visualization and analysis of atomistic simulation data with
  ovito–the open visualization tool.
\newblock \emph{Modelling and Simulation in Materials Science and Engineering},
  18\penalty0 (1):\penalty0 015012, December 2009.
\newblock ISSN 1361-651X.
\newblock \doi{10.1088/0965-0393/18/1/015012}.
\newblock URL \url{http://dx.doi.org/10.1088/0965-0393/18/1/015012}.

\bibitem[Frank and Kasper(1959)]{frankComplexAlloyStructures1959}
F.~C. Frank and J.~S. Kasper.
\newblock Complex alloy structures regarded as sphere packings. {{II}}.
  {{Analysis}} and classification of representative structures.
\newblock \emph{Acta Crystallographica}, 12\penalty0 (7):\penalty0 483--499,
  July 1959.
\newblock ISSN 0365-110X.
\newblock \doi{10.1107/s0365110x59001499}.

\bibitem[Goryaeva et~al.(2023)Goryaeva, Domain, Chartier, Dézaphie, Swinburne,
  Ma, Loyer-Prost, Creuze, and Marinica]{goryaevaCompactA15FrankKasper2023}
Alexandra~M. Goryaeva, Christophe Domain, Alain Chartier, Alexandre Dézaphie,
  Thomas~D. Swinburne, Kan Ma, Marie Loyer-Prost, Jérôme Creuze, and
  Mihai-Cosmin Marinica.
\newblock Compact a15 frank-kasper nano-phases at the origin of dislocation
  loops in face-centred cubic metals.
\newblock \emph{Nature Communications}, 14\penalty0 (1), May 2023.
\newblock ISSN 2041-1723.
\newblock \doi{10.1038/s41467-023-38729-6}.
\newblock URL \url{http://dx.doi.org/10.1038/s41467-023-38729-6}.

\bibitem[Wyszkowska et~al.(2025)Wyszkowska, Mieszczyński, Aligayev, Azarov,
  Chromiński, Kalita, Kosińska, Dominguez-Gutierrez, Kurpaska, Jóźwik, and
  Jagielski]{wyszkowskaNanoscaleDefectFormation2025}
Edyta Wyszkowska, Cyprian Mieszczyński, Amil Aligayev, Alexander Azarov,
  Witold Chromiński, Damian Kalita, Anna Kosińska, Francisco~Javier
  Dominguez-Gutierrez, Łukasz Kurpaska, Iwona Jóźwik, and Jacek Jagielski.
\newblock Nanoscale defect formation in fcc ni and low-fe content
  ni$_x$fe$_{1-x}$ single crystals induced by self-ion irradiation.
\newblock \emph{Nanoscale}, 17\penalty0 (26):\penalty0 15841–15855, 2025.
\newblock ISSN 2040-3372.
\newblock \doi{10.1039/d5nr00117j}.
\newblock URL \url{http://dx.doi.org/10.1039/d5nr00117j}.

\bibitem[Del~Fr{\'e} et~al.(2025)Del~Fr{\'e}, De~Backer, Domain, Thuinet, and
  Becquart]{NiPd_dataset}
Samuel Del~Fr{\'e}, Andr{\'e}e De~Backer, Christophe Domain, Ludovic Thuinet,
  and Charlotte~S. Becquart.
\newblock Cascades\_ni\_fenicr\_zr, 2025.
\newblock URL \url{https://doi.org/10.17632/x543g8r5sz.1}.

\end{thebibliography}

\end{document}